\documentclass[%
 aip,
 reprint,%
]{revtex4-1}

\usepackage{lipsum}
\usepackage{graphicx}
\usepackage{dcolumn}
\usepackage{bm}
\usepackage{tabularx, multirow, booktabs}
\usepackage{xcolor}
\usepackage{comment}

\usepackage[utf8]{inputenc}
\usepackage[T1]{fontenc}
\usepackage{newtxtext}
\usepackage{amsmath}
\usepackage[varvw]{newtxmath}
\newcommand\Alfven{Alfv\'en }
\newcommand\Alfvenic{Alfv\'enic }

\newcommand{\eqr}[1]{Eq.\thinspace(#1)}
\newcommand{\pfrac}[2]{\frac{\partial #1}{\partial #2}}

\newcommand{\mvec}[1]{\mathbf{#1}}
\newcommand{\gvec}[1]{\boldsymbol{#1}}

\newcommand{\tbasis}[1]{\tilde{\mvec{e}}_{#1}}
\newcommand{\tdbasis}[1]{\tilde{\mvec{e}}^{#1}}

\newcommand{\delx}{\nabla_\mvec{x}}

\newcommand{\delxpp}{\nabla_{\mvec{x}''}}

\newcommand{\bdf}{\bar{f}}

\newcommand{\buni}{\mvec{b}}

\newcommand{\gke}{\texttt{Gkeyll}}

\draft 

\begin{document}

\title[The Kinetic Pressure-Strain]{The Kinetic Analogue of the Pressure-Strain Interaction}

\author{S.~A.~Conley}
\affiliation{Department of Astrophysical Sciences, Princeton University, Princeton, NJ 08544, USA.}
\email[E-mail: ]{sarah.conley@princeton.edu}
\author{J.~Juno} 
\affiliation{Princeton Plasma Physics Laboratory, Princeton, NJ 08540, USA.}
\author{J.~M.~TenBarge}
\affiliation{Department of Astrophysical Sciences, Princeton University, Princeton, NJ 08544, USA.}
\author{M.~H.~Barbhuiya}
\author{P.~A.~Cassak}
\affiliation{Department of Physics and Astronomy and the Center for KINETIC Plasma Physics, West Virginia University, Morgantown, WV 26506, USA.}
\author{G.~G.~Howes}
\affiliation{Department of Physics and Astronomy, University of Iowa, Iowa City, IA 52242, USA.}
\author{E.~Lichko}
\affiliation{Department of Astronomy and Astrophysics, University of Chicago, Chicago, IL 60637 USA.}
\date{\today}

\begin{abstract}

Energy transport in weakly collisional plasma systems is often studied with fluid models and diagnostics. However, the applicability of fluid models is necessarily limited when collisions are weak or absent, and using a fluid approach can obscure kinetic processes that provide key insights into the physics of energy transport. A kinetic technique that retains all of the information in 3D-3V phase-space for the study of energy transfer between electromagnetic fields and particle kinetic energy, which is quantified by the rate of electromagnetic work per unit volume $\mathbf{j}\cdot\mathbf{E}$ in fluid models, is the Field-Particle Correlation (FPC) technique. This technique has demonstrated that leveraging the full information contained in phase-space can elucidate the physical mechanisms of energy transfer. This provides a significant advantage over fluid diagnostics that quantify the rate at which energy is exchanged but do not distinguish between different physical processes. A different channel of energy transport, between fluid flow energy and particle internal energy, is quantified in fluid models via the pressure-strain interaction $-(\mathbf{P} \cdot \nabla) \cdot \mathbf{u}$. Using a similar approach to that of the field-particle correlation technique, in this work we derive a kinetic analog of the pressure-strain interaction and use it alongside the field-particle correlation to analyze the flow of energy from electromagnetic fields into particle internal energy in two case studies of electron Landau damping.

\end{abstract}

\maketitle
\section{Introduction}
\label{sec:Intro}  

A major goal in heliophysics, planetary sciences, and astrophysics is to develop a clear understanding of energy transport in collisionless plasmas. This is key to fully address a variety of open questions regarding weakly collisional plasma systems, including the heating of the solar corona\citep{Cranmer:2015}, dissipation of turbulence in the solar wind\citep{Coleman:1968} and interstellar medium\citep{Minter:1997}, planetary magnetosphere dynamics\citep{Chaston:2008}, and energy transfer in magnetic reconnection\citep{Aunai:2011}. In pursuit of this goal, energy transport diagnostics are often used to identify regions where energy is being converted between different forms in a plasma system. Several of these diagnostics are derived from the Vlasov equation, which fully describes the kinetic evolution of a collisionless plasma system when coupled with Maxwell's equations for the electromagnetic fields. Formally, energy transport in this model is reversible since collisions are neglected in the Vlasov equation. For a physical system, however, it is usually assumed that weak collisions will ultimately thermalize the energy transferred to the particles, yielding irreversible heating of the plasma\citep{Schekochihin:2009}.

Due to its complexity and computational and diagnostic expense, the full six-dimensional plus time phase-space plasma description contained in the Vlasov equation is often sacrificed for more tractable fluid variables. This is done without making any immediate assumptions about the system by calculating the velocity moments of the Vlasov equation to produce a hierarchy of three-dimensional fluid equations. The first two moments of the Vlasov equation yield evolution equations for the particle density and momentum. The third (scalar) moment generates an evolution equation for the particle kinetic energy density $\mathcal{E}^k$, which is used to derive various diagnostics of energy transport. Namely, for infinite or periodic boundary conditions, the rate of electromagnetic work on the particles per unit volume $\mathbf{j}\cdot\mathbf{E}$, where $\mathbf{j}$ is the current density and $\mathbf{E}$ is the electric field, is responsible for net transfer between electromagnetic field energy and particle kinetic energy\citep{Klein:2016,Howes:2017}. Motivated by this, $\mathbf{j}\cdot\mathbf{E}$ has been heavily studied as a diagnostic of energy transfer in collisionless plasma systems\citep{Zenitani:2011,Wan:2015}. However, though $\mathbf{j}\cdot\mathbf{E}$ quantifies the rate of energy density transfer from fields to particles at a given location, like all fluid diagnostics, it cannot be used to ascertain the physical mechanism driving the energy transfer. A method of studying energy transfer described by $\mathbf{j}\cdot\mathbf{E}$ while retaining full phase-space information is known as the Field-Particle Correlation (FPC) technique\citep{Klein:2016,Howes:2017,Klein:2017}. Field-particle correlations have been used to study collisionless plasma energization in simulations\citep{Juno:2018,Montag:2022,Conley:2023,Huang:2024}, spacecraft observations\citep{Chen:2019,Afshari:2021}, and laboratory experiments\citep{Schroeder:2021}, and have revealed the immense value of retaining all of phase-space in the study of energy transfer. Specifically, this technique has shown that different energization mechanisms produce unique signatures in velocity space which provide insight into the kinetic physics responsible for energy transfer at a given location\citep{Juno:2018,Chen:2019,Montag:2022,Huang:2024}.

Another energy transport diagnostic is derived by separating the kinetic energy density of a collisionless system into internal energy density of the particles (the rest-frame kinetic energy density) $\mathcal{E}^{int}$, and energy density of the bulk flow $\mathcal{E}^f$, such that $\mathcal{E}^k = \mathcal{E}^{int} + \mathcal{E}^f$. It was recently emphasized that, when kinetic energy density is divided in this way and the separate evolution equations for the internal and fluid flow energy densities are spatially averaged over infinite or periodic boundary conditions, $\mathbf{j}\cdot\mathbf{E}$ is a source term for the time evolution of $\mathcal{E}^f$ only, and therefore quantifies energy transfer between the electromagnetic fields and the kinetic energy of the bulk flow. A term called the pressure-strain interaction, $-(\mathbf{P}\cdot\nabla)\cdot \mathbf{u}$, is analogously the source/sink term for conversion between fluid flow energy and internal energy \citep{Yang:2017a,Yang:2017b}, where $\mathbf{P}$ is the pressure tensor and $\mathbf{u}$ is the bulk flow velocity. Since the role of the pressure-strain interaction in describing energy conversion between bulk flow and internal energy was brought to the attention of the community, it has been extensively used to analyse both simulation data\citep{Sitnov:2018,Parashar:2018,Pezzi:2019} and spacecraft observations \citep{Chasapis:2018,Bandyopadhyay:2018,Zhong:2019}. Like $\mathbf{j}\cdot\mathbf{E}$, the pressure-strain interaction measures a rate of change of energy density but obscures the kinetic details within velocity space that enable deeper understanding of energy transport.

In this work, we introduce a new phase-space diagnostic tool that accomplishes for energy conversion via the pressure-strain interaction what the the field-particle correlation technique accomplishes for energy transfer via the electromagnetic work: the  \textit{
Kinetic Pressure-Strain} (KPS). This diagnostic retains full phase-space information for the term that gives rise to the pressure-strain interaction, and thus reveals which particles are participating in the exchange of energy described by $-(\mathbf{P}\cdot\nabla)\cdot\mathbf{u}$ as a function of velocity. The rest of this paper is organized as follows: in Sec.~\ref{sec:rev} we review details of the dissipation measures that we employ; in Sec.~\ref{sec:kpw}, we derive the kinetic pressure-strain; in Sec.~\ref{sec:analysis}, we describe our simulations of (1) a standing Langmuir wave and (2) a traveling \Alfven wave; in Sec.~\ref{sec:kps_analysis} we use these as case studies for diagnosing energy conversion using kinetic pressure-strain, and provide comparisons of the KPS with the fluid quantity $-(\mathbf{P} \cdot \nabla) \cdot \mathbf{u}$ and with field-particle correlation signatures; and in Sec.~\ref{sec:disc} we summarize and conclude.

\section{Review of Select Dissipation Measures}
\label{sec:rev}

Taking the second moment of the Vlasov equation by multiplying by $(1/2) m_s \mathbf{v}^2$, where $m_s$ is the species mass and $\mathbf{v}$ is the velocity coordinate, and integrating over all velocity space yields an expression for the rate of change of the total kinetic energy density of a plasma species $s$, $\mathcal{E}^{k}_s = \int d^3v\ (1/2) m_s \mathbf{v}^2 f_s$, where $f_s(\mathbf{x}, \mathbf{v}, t)$ is the six-dimensional distribution function:
\begin{equation}
    \frac{\partial {\mathcal{E}^{k}_s}}{\partial t} + \nabla \cdot (\mathcal{E}^{k}_s \mathbf{u}_s) = -\nabla \cdot (\mathbf{P}_s \cdot \mathbf{u}_s) - \nabla \cdot \mathbf{q}_s + \mathbf{j}_s \cdot \mathbf{E},
    \label{eq:kin_e}
\end{equation}
where $\mathbf{u}_s = (1/n_s)\int d^3v\ \mathbf{v} f_s$ is the fluid velocity, $n_s = \int d^3v f_s$ is the particle density, $\mathbf{P}_s = m_s \int d^3v\ (\mathbf{v} - \mathbf{u}_s)(\mathbf{v} - \mathbf{u}_s) f_s$ is the pressure tensor, $\mathbf{q}_s = (m_s/2)\int d^3v\ |\mathbf{v} - \mathbf{u}_s|^2(\mathbf{v}-\mathbf{u}_s) f_s$ is the heat flux density vector, $\mathbf{j}_s = q_s n_s \mathbf{u}_s$ is the species current density, and $q_s$ is the particle charge. After spatially integrating this equation over infinite or periodic boundary conditions, $\mathbf{j}_s\cdot\mathbf{E}$ is the only source term for $dE_s^k/dt$, where $E_s^k=\int d^3x\ \mathcal{E}_s^k$ is the kinetic energy. The remaining are transport terms: these terms may be very important in the time rate of change of the local particle kinetic energy density \citep{Pezzi:2019}, but when integrated over an entire spatial domain they simply move energy from one location to another or across a boundary. 

The field-particle correlation technique describes energy density transfer by $\mathbf{j}_s\cdot\mathbf{E}$ in full phase-space. FPCs are evaluated at a single point in physical space $\mathbf{x}_0$ and are often integrated over a period of time $\tau$ to remove oscillatory components of the energy transfer. Though different components of the electromagnetic fields may be chosen depending on the physics of interest\citep{Juno:2023,Huang:2024}, for isolating energization due to the parallel electric field during the process of Landau damping, the standard form of the FPC is\citep{Klein:2016,Howes:2017,Klein:2017,Chen:2019}
\begin{equation}
    C_{E_\parallel}(\mathbf{v},t) = -q_s \frac{v_\parallel^2}{2}\frac{\partial f_s(\mathbf{x}_0,\mathbf{v},t)}{\partial v_\parallel} E_\parallel(\mathbf{x}_0,t),
    \label{eq:fpc}
\end{equation}
where $v_\parallel$ and $E_\parallel$ are the velocity and electric field components parallel to the magnetic field $\mathbf{B}$, if one is present. Note that the factor of $v_\parallel^2$ appears in this form of the FPC rather than $v^2 = v_\parallel^2 + v_\perp^2$ since the contribution from $v_\perp^2$ goes to zero upon integration over all of velocity space, and therefore does not contribute to the net transfer of energy\citep{Klein:2017,Juno:2021}. In scenarios where evaluating the velocity derivative in Eq.~(\ref{eq:fpc}) leads to large errors, an alternative form of the field-particle correlation may be used instead\citep{Klein:2017},
\begin{equation}
    C^\prime_{E_\parallel}(\mathbf{v},t) = q_s v_\parallel f_s(\mathbf{x_0},\mathbf{v},t)E_\parallel(\mathbf{x}_0,t).
    \label{eq:afpc}
\end{equation}
Both $C_{E_\parallel}$ and $C_{E_\parallel}^\prime$ are equal to the fluid quantity $j_{\parallel s} E_\parallel$ when integrated over velocity. Physically, the standard form of the field-particle correlation measures the total rate of change of phase-space kinetic energy density, $w_s = (1/2)m_s |\mathbf{v}|^2 f_s$, where $\int d^3v\ w_s = \mathcal{E}^k_s$, at a given point in $(\mathbf{x},\mathbf{v})$ space\citep{Klein:2016,Howes:2017}, while the alternative form omits a term describing a flux of phase-space energy density and so contains only the local rate of change of $w_s$ that contributes directly to $\mathbf{j}\cdot\mathbf{E}$.

A plasma species' total kinetic energy density can be separated into $\mathcal{E}^k_s = \mathcal{E}^f_s + \mathcal{E}^{int}_s$, where $\mathcal{E}^{f}_s = (1/2)m_s n_s |\mathbf{u}_s|^2$ is the fluid flow energy density, and $\mathcal{E}^{int}_s  = \int d^3v\ (1/2) m_s |\mathbf{v} - \mathbf{u}_s|^2 f_s$ is the internal energy density (sometimes referred to as the thermal energy density). The evolution equations for these, coupled with the evolution of the electromagnetic energy density $\mathcal{E}^m = (1/2) \left( \mathbf{B}^2 / \mu_0 + \epsilon_0 \mathbf{E}^2\right)$, are
\begin{equation}
    \frac{\partial \mathcal{E}^f_s}{\partial t} + \nabla \cdot (\mathcal{E}^f_s \mathbf{u}_s )= -\nabla \cdot (\mathbf{P}_s \cdot \mathbf{u}_s) + (\mathbf{P}_s \cdot \nabla) \cdot \mathbf{u}_s + \mathbf{j}_s \cdot \mathbf{E},
    \label{eq:f_e}
\end{equation}
\begin{equation}
    \frac{\partial \mathcal{E}^{int}_s}{\partial t} + \nabla \cdot (\mathcal{E}^{int}_s \mathbf{u}_s) = -(\mathbf{P}_s \cdot \nabla) \cdot \mathbf{u}_s - \nabla \cdot \mathbf{q}_s,
    \label{eq:int_e}
\end{equation}
\begin{equation}
    \frac{\partial \mathcal{E}^m}{\partial t} + \frac{1}{\mu_0} \nabla \cdot (\mathbf{E} \times \mathbf{B}) = -\mathbf{j} \cdot \mathbf{E},
    \label{eq:em_e}
\end{equation}
where $\mathbf{j} = \sum_s \mathbf{j}_s$. Though these equations are well-known, it was recently emphasized that after taking a spatial integral over infinite or periodic boundary conditions,  $\mathbf{j}_s\cdot\mathbf{E}$ and $-(\mathbf{P}_s\cdot\nabla)\cdot\mathbf{u}_s$ are the only terms driving net changes in the energy densities, since the divergence terms integrate to zero. Therefore, these two terms reveal the pathways through which energy must be converted globally between its different fluid forms in infinite or periodic systems \citep{Yang:2017a,Yang:2017b}. Namely, the rate of electromagnetic work, $\int d^3x\ \mathbf{j} \cdot \mathbf{E}$, is a source for only $dE^m/dt$ and $dE_s^f/dt$, where $E^m = \int d^3x\ \mathcal{E}^m$ and $E_s^f = \int d^3x\ \mathcal{E}_s^f$, and therefore is responsible for net transfer between the electromagnetic field and fluid flow energies. The spatially-integrated pressure-strain interaction, $\int d^3x\ \left[-(\mathbf{P}_s \cdot \nabla) \cdot \mathbf{u}_s\right]$, is a source for only $dE_s^f/dt$ and $dE_s^{int}/dt$, where $E_s^{int} = \int d^3x\ \mathcal{E}_s^{int}$, and therefore mediates conversion between fluid flow and internal energies. Note that throughout this paper, we use the term \textit{transfer} to refer to energy transport between fields and particle kinetic energy and \textit{conversion} to refer to energy transport between different degrees of freedom within particle kinetic energy.

To elucidate the physics of the pressure-strain interaction, it is often decomposed into the pressure dilatation, $-\mathcal{P}_s \theta_s$, and $-\mathbf{\Pi}_s:\mathbf{D}_s$ (called Pi-D)\citep{Yang:2017a},
\begin{equation}
    -(\mathbf{P}_s \cdot \nabla) \cdot \mathbf{u}_s = -\mathcal{P}_s \theta_s - \mathbf{\Pi}_s : \mathbf{D}_s,
    \label{eq:pdivu_PiD}
\end{equation}
where $\mathcal{P}_s = (1/3) P_{s,ii}$ is the scalar pressure (with the Einstein summation convention implied on repeated indices), $\theta_s=(1/3)\partial_i u_{s,i}$ is the dilatation,  $\mathbf{\Pi_s} = \mathbf{P}_s - \mathcal{P}_s \mathbf{I}$ is the deviatoric pressure tensor, and $\mathbf{I}$ is the identity tensor. The traceless strain-rate tensor is $\mathbf{D}_{s} = \mathbf{S}_s - \theta_s \mathbf{I}$, where $S_{s, ij} = (1/2) \left( \partial_i u_{s,j} + \partial_j u_{s,i} \right)$ is the strain-rate tensor. In this decomposition, the pressure dilatation describes isotropic volume compression/expansion and Pi-D describes volume-preserving deformation\citep{DelSarto:2018}. 

\section{The Kinetic Pressure-Strain Interaction}\label{sec:kpw}

Here, we analyze the pressure-strain interaction in a novel way by considering the term that gives rise to $-(\mathbf{P}\cdot\nabla)\cdot\mathbf{u}$ in full phase-space. This kinetic version of the pressure-strain interaction is straightforward to derive by starting with the Vlasov equation in the plasma rest frame\citep{Ramos:2008} (but see Appendix~\ref{sec:labframe} for a derivation starting from the stationary frame),
\begin{multline}
    \frac{\partial f_s}{\partial t} +  \nabla \cdot \left[\left(\mathbf{v}^\prime + \mathbf{u}_s\right) f_s \right] + \nabla_{v^\prime} \cdot \left\{ \left[ \frac{q_s}{m_s} (\mathbf{E} + \mathbf{u}_s \times \mathbf{B} \right. \right. \\
     \left. \left.  + \mathbf{v}^\prime \times \mathbf{B}) - \frac{\partial \mathbf{u}_s}{\partial t} - \mathbf{u}_s \cdot \nabla\mathbf{u}_s - \mathbf{v}^\prime \cdot \nabla\mathbf{u}_s \right] f_s \right\}=0,
    \label{eq:rfV}
\end{multline}
where the coordinate transformation of the velocity is $\mathbf{v}^\prime = \mathbf{v} - \mathbf{u}_s(\mathbf{x}, t)$, but $\mathbf{x}$ and $t$ are untransformed. Note that the transformed velocity coordinate differs for each species, but we omit the species subscript here for consistency. Next, the equations for mass and momentum conservation are used to generate an expression for the rate of change of the fluid velocity,
\begin{equation}
    \frac{\partial \mathbf{u}_s}{\partial t} + \mathbf{u}_s \cdot \nabla \mathbf{u}_s + \frac{1}{\rho_s} \nabla \cdot \mathbf{P}_s = \frac{q_s}{m_s}(\mathbf{E} + \mathbf{u}_s \times \mathbf{B}),
\end{equation}
which is substituted into Eq.~(\ref{eq:rfV}) to yield
\begin{multline*}
    \frac{\partial f_s}{\partial t} + \nabla \cdot \left[\left(\mathbf{v}^\prime + \mathbf{u}_s\right) f_s \right] + \\
    \nabla_{v^\prime} \cdot  \left[ \left( \frac{1}{\rho_s} \nabla \cdot \mathbf{P}_s\ + \frac{q_s}{m_s} \mathbf{v}^\prime \times \mathbf{B} - \mathbf{v}^\prime \cdot \nabla\mathbf{u}_s \right) f_s \right] = 0. \\
\end{multline*}
By multiplying this form of the Vlasov equation by $\frac{1}{2} m_s |\mathbf{v}^{\prime}|^2$ and defining $ w^\prime_s(\mathbf{x},\mathbf{v}^\prime,t)=\frac{1}{2} m_s|\mathbf{v}^{\prime}|^2 f_s(\mathbf{x},\mathbf{v}^\prime,t)$ as the phase-space internal energy density, we get an equation for the rate of change of $w^\prime_s$ 
\begin{equation}
\begin{split}
\frac{\partial w^\prime_s}{\partial t} & + \nabla \cdot \left[\left(\mathbf{v}^\prime + \mathbf{u}_s \right) w^\prime_s  \right] + \frac{1}{2} m_s|\mathbf{v}^\prime|^2 \nabla_{v^\prime} \cdot \\ & \left[ \left( \frac{1}{\rho_s} \nabla \cdot \mathbf{P}_s\ + \frac{q_s}{m_s} \mathbf{v}^\prime \times \mathbf{B} - \mathbf{v}^\prime \cdot \nabla\mathbf{u}_s \right) f_s \right]  = 0,
\label{eq:rf_dEintdt1}
\end{split}
\end{equation}
where partial derivatives in space and time are evaluated at constant $\mathbf{v}^\prime$, which allows the transformed velocity coordinate to pass through them\citep{Thorne:2017}. 

When integrated over all of velocity space, Eq.~(\ref{eq:rf_dEintdt1}) describes the evolution of the internal energy density $\mathcal{E}^{int}_s$, which is $\mathcal{E}^{int}_s = \int d^3v^\prime (1/2) m_s |\mathbf{v}^\prime|^2 f_s$ in terms of $\mathbf{v}^\prime$. Therefore, the velocity integrals of the first two terms are straightforward: $$\int d^3v^\prime\ \frac{\partial w_s^\prime}{\partial t} = \frac{\partial \mathcal{E}^{int}_s}{\partial t},$$
and $$\int d^3v^\prime\ \nabla \cdot \left[\left( \mathbf{v}^\prime + \mathbf{u}_s \right) w_s^\prime \right] = \nabla \cdot \left( \mathbf{q}_s + \mathcal{E}^{int}_s \mathbf{u}_s \right).$$
The third term of Eq.~(\ref{eq:rf_dEintdt1}) has three pieces which can each be evaluated through integration by parts. These become
$$-\frac{1}{n_s}\nabla \cdot \mathbf{P}_s \int d^3v^\prime\ f_s \mathbf{v}^\prime = 0,$$
since there is zero flow in the plasma rest frame;
$$- \frac{q_s}{2} \int d^3v^\prime f_s \left(\mathbf{v}^\prime \times \mathbf{B} \right) \cdot \mathbf{v}^\prime = 0,$$
upon using a vector identity; and
$$m_s \int d^3v^\prime f_s \left( \mathbf{v}^\prime\mathbf{v}^\prime \cdot \nabla \right) \cdot \mathbf{u}_s = \left( \mathbf{P}_s \cdot \nabla \right) \cdot \mathbf{u}_s.$$
Therefore, we obtain
$$\frac{\partial \mathcal{E}^{int}_s}{\partial t} + \nabla \cdot \left( \mathbf{q}_s + \mathcal{E}^{int}_s \mathbf{u}_s \right) + \left( \mathbf{P}_s \cdot \nabla \right) \cdot \mathbf{u}_s = 0,$$
which is equivalent to Eq.~(\ref{eq:int_e}), and from which we see that the final term in Eq.~(\ref{eq:rf_dEintdt1}) gives rise to the pressure-strain interaction. This implies that this final term is the desired form for the \textit{Kinetic Pressure-Strain} (KPS). 

Specifically, by direct analogy to the field-particle correlation we define here two complementary kinetic forms of the pressure-strain interaction that appear in the previous derivation: (i) the \emph{Kinetic Pressure-Strain}, related to the original form of the final term in Eq.~(\ref{eq:rf_dEintdt1}),
\begin{equation}
    K_{PS} \equiv \frac{m_s}{2} |\mathbf{v}^\prime|^2 \left\{ \nabla_{\mathbf{v}^\prime} \cdot \left[\left(\mathbf{v}^\prime \cdot \nabla \mathbf{u}_s\right)f_s \right] \right\};
    \label{eq:kpw}
\end{equation}
and (ii) the \emph{Alternative Kinetic Pressure-Strain}, which integrates to the same result as the kinetic pressure-strain, but does not include a velocity space flux that integrates to zero,
\begin{equation}
    \widetilde{K}_{PS} \equiv  -m_s f_s (\mathbf{v}^\prime \mathbf{v}^\prime \cdot \nabla) \cdot \mathbf{u}_s.
    \label{eq:akpw}
\end{equation}
Note that we include a negative sign in both forms for consistency with the standard definition of the pressure-strain interaction, $-(\mathbf{P}_s \cdot \nabla)\cdot \mathbf{u}_s$, and we use a tilde to differentiate the second form from the first. 

Both forms of the KPS measure rates of change of phase-space internal energy density $w_s^\prime$. The alternative kinetic 
pressure-strain in Eq.~(\ref{eq:akpw}), $\widetilde{K}_{PS}$, integrates directly to the fluid pressure-strain interaction. It measures only changes in $w_s^\prime$ that contribute to $-(\mathbf{P}_s \cdot \nabla)\cdot\mathbf{u}_s$ and therefore to changes in $\mathcal{E}^{int}_s$. The phase-space structure of the kinetic pressure-strain in Eq.~(\ref{eq:kpw}), $K_{PS}$, contains all of the structure in $\widetilde{K}_{PS}$ plus phase-space structure due to convection of $w_s^\prime$ in velocity space.  When integrated over $\mathbf{v}^\prime$, this second contribution yields zero net energy density change and therefore does not contribute to the fluid pressure-strain interaction or to changes in internal energy density, but it does measure the net change in $w_s^\prime$ at a given point in ($\mathbf{x},\mathbf{v}^\prime$) phase-space. Therefore, $K_{PS}$ measures the total change in phase-space energy density $w_s^\prime$ that arises from the relevant term in Eq.~(\ref{eq:rf_dEintdt1}), while $\widetilde{K}_{PS}$ isolates changes in $w_s^\prime$ that directly contribute to the fluid pressure-strain interaction.   

Just as for the standard and alternative forms of the field-particle correlation---which are equivalent when integrated over velocity\citep{Howes:2017}---both forms of the KPS are valuable, but depending on the scenario, one form may be more useful than the other. For instance, the standard FPC (as in Eq.~(\ref{eq:fpc})) is typically used to study Landau damping in simulations\citep{Klein:2017,Conley:2023}, since the inclusion of the term describing flux of $w_s$ in velocity space in this form leads to distinct, bipolar resonant structures with zero-crossings corresponding to the phase-velocity $v_{ph}$ of the damped wave. This flux of energy density from below to above the resonant velocity directly corresponds to acceleration of resonant particles by the parallel electric field in the process of Landau damping\citep{Horvath:2020}.  The resonant structure in the alternative FPC (as in Eq.~(\ref{eq:afpc})) has a peak that is roughly centered at $v_{ph}$, showing that the net transport of energy density occurs at the resonant velocity, but making the quantitative tie to the Landau resonance less distinct. In spacecraft observations where velocity derivatives are difficult to obtain, the alternative FPC is easier to calculate directly, although often this measure can be differentiated appropriately to yield the standard FPC \citep{Chen:2019,Afshari:2021}. In the simulations performed for this study, we find similar phase-space structures for Landau damping in $K_{PS}$ (bipolar resonant structures with zero-crossing at $v_{ph}$) and $\widetilde{K}_{PS}$ (resonant signatures peaking at $v_{ph}$). Since $\widetilde{K}_{PS}$ is analogous to the alternative form of the FPC, we refer to it as the alternative form of the KPS. We opt to use $K_{PS}$ to take advantage of its distinct, bipolar Landau-resonant structures in the majority of the following analysis, but also use $\widetilde{K}_{PS}$ in one application for its direct correspondence to $-(\mathbf{P}_s\cdot\nabla)\cdot \mathbf{u}_s$. 

\section{Simulations} \label{sec:analysis}
\subsection{Code Description}

We investigate the kinetic pressure-strain by considering two case studies of electron Landau damping: Landau damping of (1) a standing Langmuir wave and (2) a propagating \Alfven wave. Both are simulated using \gke, a discontinuous Galerkin (DG) finite element code that contains a variety of solvers relevant to plasma physics\citep{Hakim:2006,Hakim:2008,Shi:2015,Juno:2018}. For this work, we employ the Vlasov-Maxwell model within \gke\ with a newly implemented hybrid discretization that separates the dynamics parallel and perpendicular to the local magnetic field. The parallel dynamics are treated with the DG solver discussed in Juno et al.~2018\citep{Juno:2018} and the perpendicular dynamics with a spectral expansion. This approach greatly reduces the computational expense of the simulation, while still capturing the parallel kinetic physics important in the process of Landau damping. Further discussion of the simulation model is contained in Appendix~\ref{app:model}.  

\subsection{Standing Langmuir Wave}
We first model a standing Langmuir wave, tracking the kinetic evolution of the electrons in 1D-1V and treating the ions as a charge-neutralizing background. The wave is initialized in an isotropic, Maxwellian plasma within a constant background magnetic field $B_{0x}/B_0=1$ via a perturbed distribution function, $f_e/f_{Me} = 1+\alpha \sin(kx)$, and electric field, $E_x/E_0 = (\alpha/k)\cos(kx)$. Here, $f_{Me}=\left[n_{0e}/(2\pi v_{te}^2)^{1/2}\right]e^{(-v^2/2v_{te}^2)}$ is the Maxwellian equilibrium distribution, $n_{0e}$ is the initial electron density, $k\lambda_{De} = 0.5$ is the wavenumber, and $\alpha=10^{-4}$ is the scale size of the perturbation. The electron Debye length is $\lambda_{De} = v_{te}/\omega_{pe}$, where $v_{te} = \sqrt{T_{0e}/m_e}$ is the electron thermal velocity, $T_{0e}$ is the initial electron temperature in energy units, and $\omega_{pe}=\sqrt{e^2n_{0e}/\epsilon_0 m_e}$ is the electron plasma frequency. Configuration space and velocity space are each resolved with $N=32$ cells, and have extents $L = 2\pi /k$ (with periodic boundary conditions) and $-6 \le v_x^\prime/v_{te} \le 6$, respectively. The final simulation time is $t \omega_{pe} = 20$, with an output cadence of $\Delta t \omega_{pe} = 0.2$. For these parameters, the dispersion relation for electrostatic waves in a hot plasma, $\omega^2 = \omega_{pe}^2 + 3(T_e/m_e) k^2$, predicts a frequency of $\omega = 1.32 \omega_{pe}$ for the Langmuir wave.  

\begin{figure}[t!]
\includegraphics[width=0.235\textwidth,height=0.235\textwidth,bb=25 30 345 355]{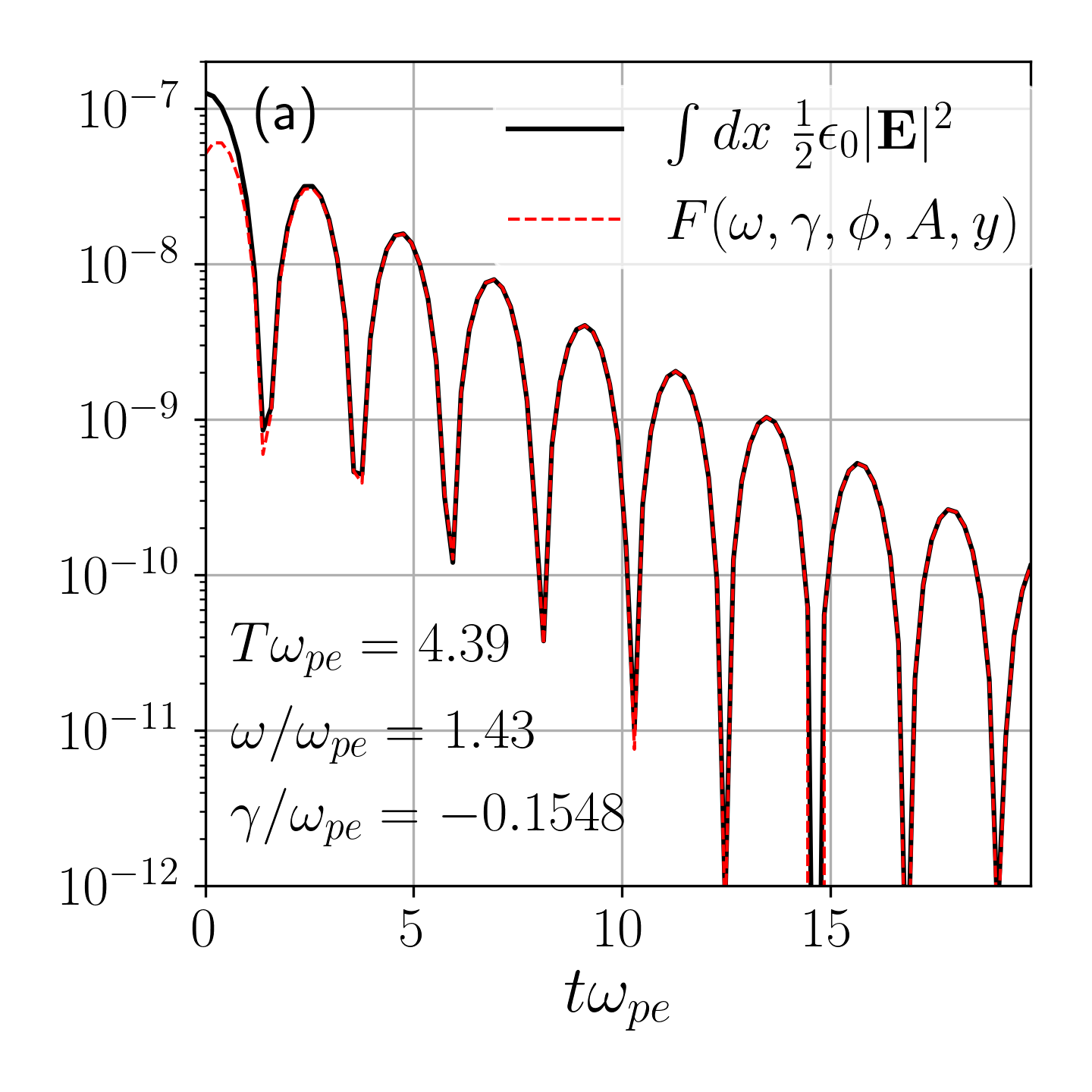}
\includegraphics[width=0.23\textwidth,height=0.235\textwidth,bb=20 30 330 338]{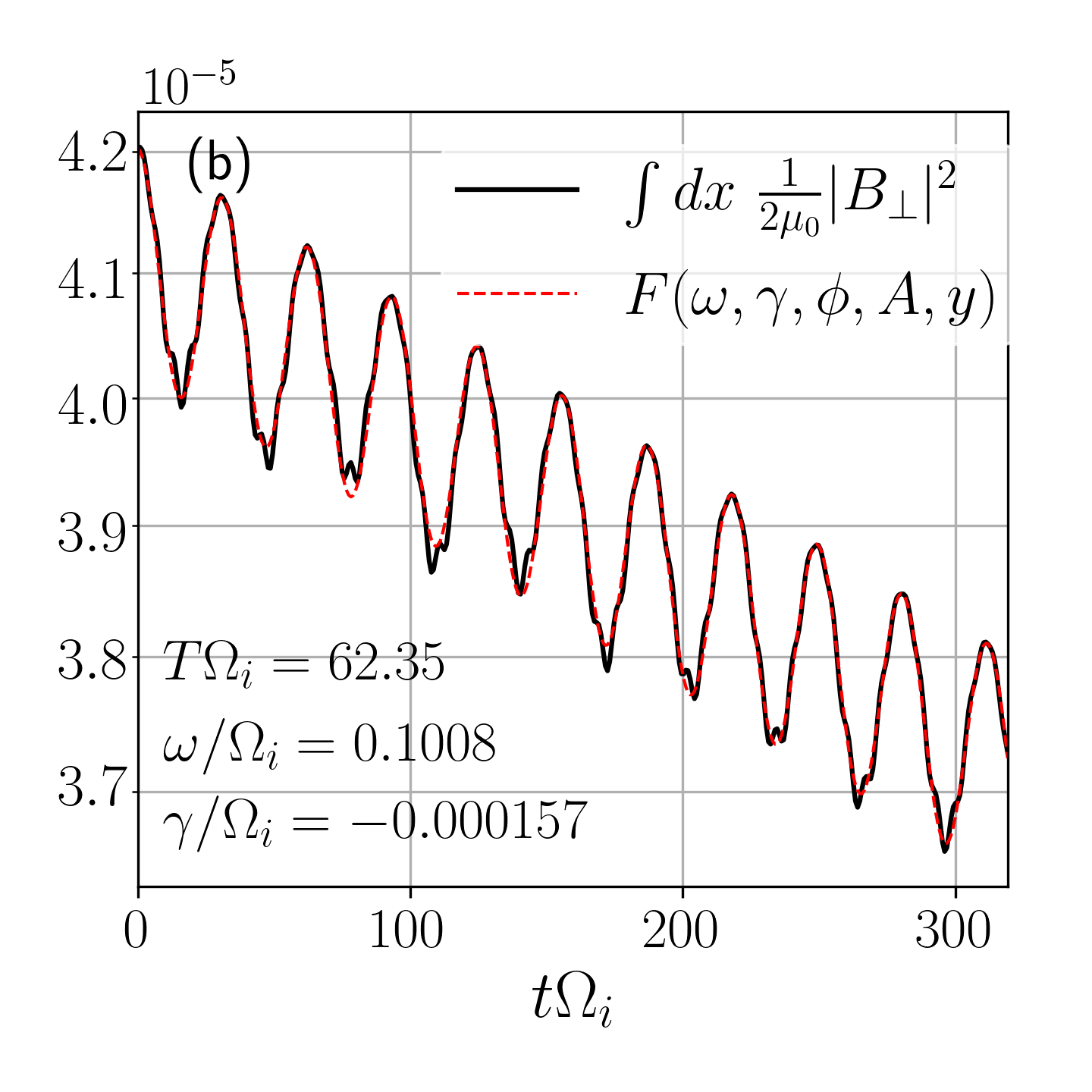}
\caption{Spatially integrated energy (black solid) (a) in the electric field of the Langmuir wave, and (b) in the perpendicular magnetic field of the \Alfven wave versus normalized time, along with lines of best fit (red dashed) given by $F(\omega, \gamma, \phi, A, y)=Ae^{2\gamma t}\left[\sin^2{\left( \omega t + \phi \right)} + y\right]$.} 
\label{fig:damping_slw}
\end{figure}

Fig.~\ref{fig:damping_slw}(a) shows the time evolution of the spatially integrated electric field energy $\int dx\ (1/2)\epsilon_0|\mathbf{E}|^2$. The field energy is fit with a function of the form, $F(\omega, \gamma, \phi, A, y)=Ae^{2\gamma t}\left[\sin^2{\left( \omega t + \phi \right)} + y \right]$, where $\omega$ is the frequency of the oscillation, $\gamma$ is the damping rate, $\phi$ the phase, $A$ the amplitude, and $y$ the vertical offset. The fit shown in (a) has $\omega = 1.43$, $\gamma = -0.1548$, $\phi = 1.04$, $A=6.78\times10^{-8}$, and $y=4.21\times10^{-4}$. This fit to the frequency and damping rate agrees very well with previous numerical results\citep{Howes:2017,Juno:2018}. The period of the Langmuir wave is calculated to be $T\omega_{pe} = 4.39$.

\begin{figure}[b!]
\includegraphics[width=0.49\textwidth,bb=15 24 465 415]{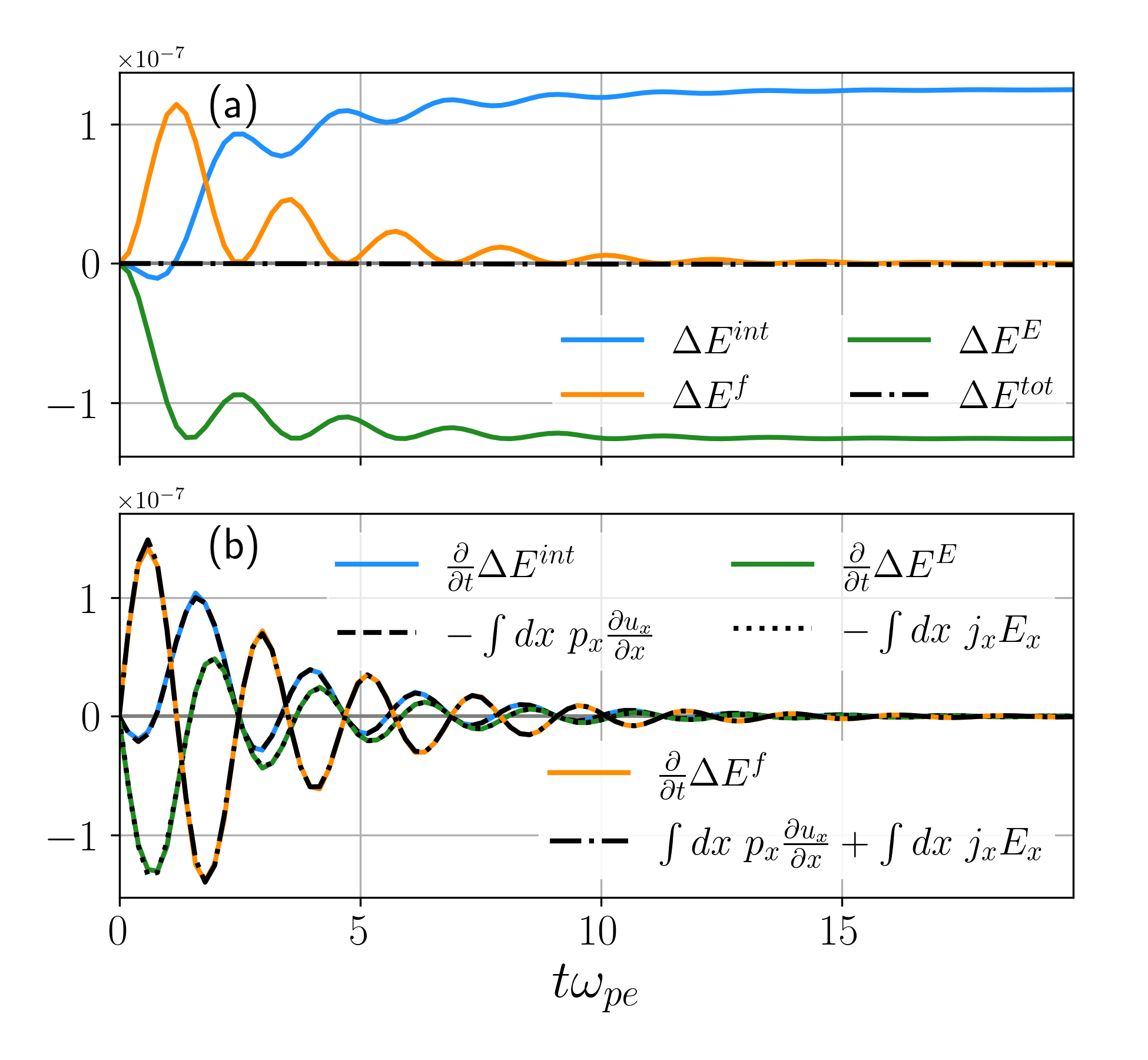}
\caption{\label{fig:energy_slw} (a) Energy balance in the Langmuir wave simulation, and (b) comparison of the time rate of change of the energy and the spatially-integrated energy conversion terms.} 
\end{figure}

The total energy $W$ in a collisionless plasma is the integral over all space of the sum of the species' internal and fluid flow energy densities, and the energy density contained in the electromagnetic fields,
\begin{equation}
\begin{split}
    W = \int d^3x\ \mathcal{E}^{tot} = \int d^3x\left[ \sum_s \left( \mathcal{E}^{int}_s + \mathcal{E}^f_s \right) + \mathcal{E}^m \right].
\end{split}
\end{equation}
The changes in energy within the Langmuir wave simulation are shown in Fig.~\ref{fig:energy_slw}(a), where $\Delta E = E - E_0$ is energy minus its initial value, and energies are obtained by integrating the energy densities over the 1D spatial domain. For the Langmuir wave, $\Delta E^m = \Delta E^E$, and species subscripts are suppressed since only electrons are dynamic in the simulation. Fig.~\ref{fig:energy_slw}(b) demonstrates that the relationships in Eqs.~(\ref{eq:f_e})-(\ref{eq:em_e}), when spatially integrated over periodic boundary conditions, hold for this simulation. Namely, changes in internal energy are governed by the box-integrated pressure-strain interaction, $\partial \Delta E^{int}/\partial t = -\int dx\ p_{x} (\partial u_{x}/\partial x)$ in 1D (blue, dashed black), changes in field energy are governed by the rate of electromagnetic work, $\partial \Delta E^E /\partial t = -\int dx\ j_{x} E_x$ in 1D (green, dotted black), and these energies may be exchanged with each other via the fluid flow energy, since $\partial \Delta E^f/\partial t  = \int dx\ \left[ p_x (\partial u_x/\partial x) +  j_x E_x \right]$ (orange, dash-dotted black). 

\subsection{Traveling \Alfven Wave}

Second, we model a traveling \Alfven wave with both kinetic electrons and ions. The wave is initialized in a gyrotropic, Maxwellian plasma with eigenfunctions calculated using the linear plasma dispersion solver \textit{PLUME}\citep{Klein:2015} for parameters $\beta_i = 2\mu_0n_0T_i/B_0^2= 0.01$, $T_i/T_e = 0.5$, and $m_i/m_e = 1836$, where $T_i$ and $T_e$ are the ion and electron temperatures, respectively. The normalized wavenumbers are $k_\parallel \rho_i = 0.01$ and $k_\perp \rho_i = 0.104$, where $\rho_i = \sqrt{2} v_{ti}/\Omega_{i}$ is the ion Larmor radius and $\Omega_{i}=eB_0/m_i$ is the ion cyclotron frequency. The parallel and perpendicular directions are defined in relation to the total, local magnetic field, which is the \Alfvenic perturbation $\delta \mathbf{B}$ plus the background field $\mathbf{B}_g = -B_0 \sin{\theta} \hat{x} + B_0 \cos{\theta} \hat{z}$, where $\theta = -\tan^{-1}(k_\parallel/k_\perp)$ is the rotation angle describing the obliquity of the wave. To enable a truly 1D-1V kinetic simulation of the \Alfven wave, the simulation is rotated so that the spatial coordinate is along the local wavevector $\mathbf{k}$ and the velocity coordinate is along the local magnetic field $\mathbf{B} = \mathbf{B}_g + \delta \mathbf{B}$. These two kinetic dimensions completely describe Landau damping due to the parallel electric field, while the perpendicular dynamics are captured using a spectral expansion (see Appendix~\ref{app:model}). Physical space has periodic boundary conditions and is resolved with $N_x = 112$ cells over $L = 2\pi/k$, where $k = \sqrt{k_\parallel^2 + k_\perp^2}$ is the magnitude of the total wavenumber. Velocity space is resolved with $N_v = 128$ cells over $-6 \le v_\parallel^\prime/v_{te} \le 6$. The simulation runs until time $t\Omega_{i} = 320$, which corresponds to $t \omega_A = 32.26$ or $t/T \approx 5$, where $T = 2 \pi/\omega_A =2 \pi/(k_\parallel v_{Ae}) $ is the wave period, with an output cadence of $\Delta t \Omega_{i} = 0.8$. The electron \Alfven speed normalized to the speed of light is $v_{Ae}/c = 0.8$. Though $v_{Ae}$ is a large fraction of $c$, the simulation dynamics remain non-relativistic since this corresponds to a maximum velocity of $v_{max} = \pm6\ v_{te} = \pm6\ v_{Ae} \sqrt{\beta_i T_e/2T_i} \approx 0.48c$ in our velocity-space extents. Note that our choice of plasma parameters is motivated by the challenge of launching an isolated \Alfven wave in a full Vlasov-Maxwell system. To accomplish this, we select a region of parameter space in which the \Alfven and fast magnetosonic modes are well separated in frequency, and in which the slow magnetosonic mode is damped more strongly than the \Alfven wave. 

Fig.~\ref{fig:damping_slw}(b) shows the time evolution of the spatially integrated perpendicular magnetic field energy $\int dx |B_\perp|^2 / 2 \mu_0$ (black) and its fit of form $F(\omega, \gamma, \phi, A, y)=Ae^{2\gamma t}\left[\sin^2(\omega t + \phi) + y \right]$ (red dashed), for which $\omega=0.1008$, $\gamma=-1.57\times10^{-4}$, $\phi=1.57$, $A=1.83\times10^{-6}$, and $y=21.98$. The period of the \Alfven wave is calculated to be $T\Omega_i=62.35$. The frequency and damping rate of the wave align well with the values predicted by the linear solver \textit{PLUME} for an \Alfven wave in these plasma parameters: $\omega/\Omega_{i} = 0.1002$ and $\gamma/\Omega_{i} = -1.58\times10^{-4}$.

\begin{figure}[b!]
\includegraphics
[width=0.49\textwidth,bb=25 24 565 560]{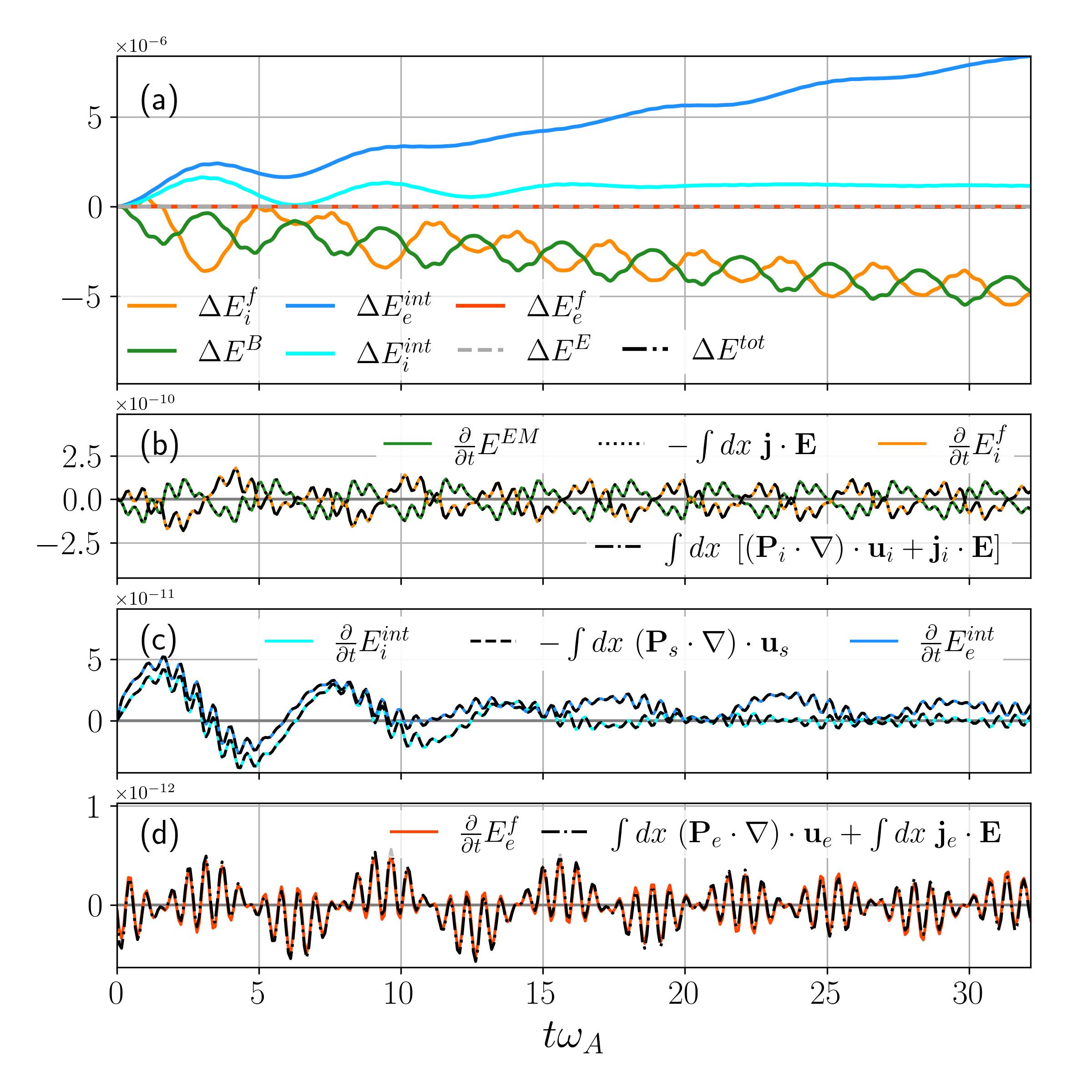}
\caption{(a) Energy conservation in the \Alfven wave simulation, and (b)-(d) balance between changes in energy and box-integrated energy transfer terms. Note that dashed black lines are plotted in (c) for both ions ($s=i$) and electrons ($s=e$); each aligns with the rate of change of that species' internal energy.}
\label{fig:aw_energybalance}
\end{figure}

\begin{figure*}[bt!]
\includegraphics[width=0.96\textwidth,bb=50 40 775 565]{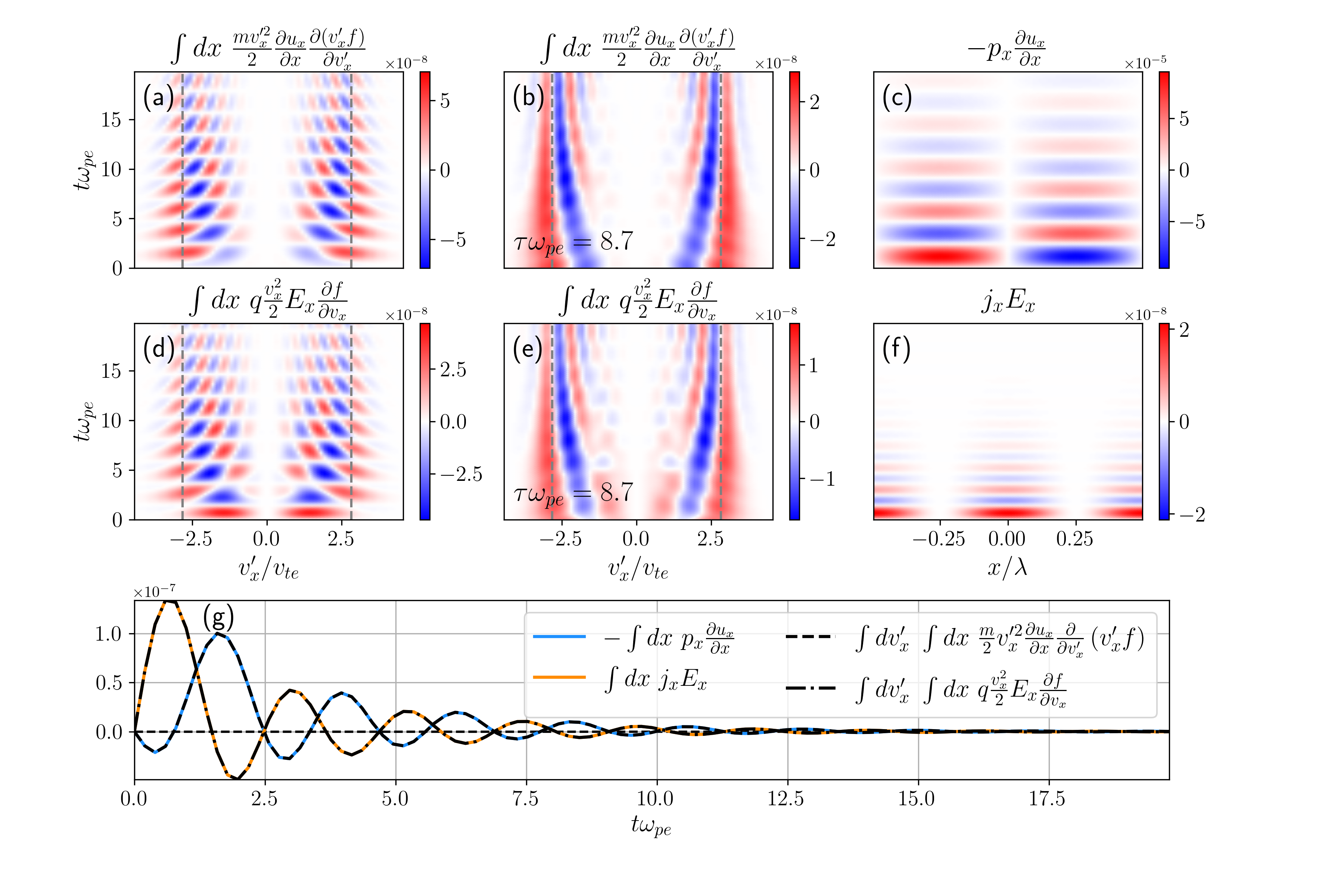}
\caption{ Comparison of the instantaneous, box-integrated (a) parallel KPS and (d) $E_\parallel$ FPC, along with their time-averaged ($\tau \omega_{pe} = 8.7$, which corresponds to approximately $2T$) values in (b) and (e), respectively, for the Langmuir wave. The (c) pressure-strain interaction and (f) rate of electromagnetic work per unit volume are also shown as functions of $x/\lambda$. Panel (g) shows (a) and (d) integrated over $v_x^\prime/v_{te}$ and (c) and (f) integrated over $x/\lambda$ as a function of time $t\omega_{pe}$. } 
\label{fig:kpw_fpc_slw}
\end{figure*}

The energy balance in the \Alfven wave simulation is shown in Fig.~\ref{fig:aw_energybalance}(a). The changes in magnetic field energy $\Delta E^B$ (green) and ion fluid flow energy $\Delta E^f_i$ (orange) show oscillatory exchange of energy at a frequency corresponding to the \Alfven wave and decay as the wave damps. The ions gain some internal energy $\Delta E^{int}_i$ (cyan) early in the simulation, and the electrons gain a larger amount of internal energy $\Delta E^{int}_e$ (blue) throughout. This is consistent with linear theory for these plasma parameters, for which the PLUME solver predicts an \Alfven wave that is primarily electron Landau damped and a low-frequency slow magnetosonic wave that is primarily ion Landau damped (with some damping onto the electrons) at a higher rate. The changes in the electric field $\Delta E^E$ (grey dashed) and the electron fluid flow $\Delta E^f_e$ (red) energies are negligible in the total box-integrated energy budget. Panels (b)-(d) compare the time derivatives of the changes in species and electromagnetic energy with their respective source terms. The terms are separated into panels according to their magnitude so that each is displayed clearly. Just as in the Langmuir wave simulation, we see good agreement between the rates of change of the energy (colored lines) and the box-integrated electromagnetic and pressure-strain terms (black dashed, dotted, or dash-dotted).

\section{Kinetic Pressure-Strain Analysis}
\label{sec:kps_analysis}

In these 1D-1V simulations, the velocity coordinate is always parallel to the local magnetic field and the kinetic pressure-strain reduces from its form in Eq.~(\ref{eq:kpw}) to the parallel kinetic pressure-strain,  
\begin{equation}
    K_{PS\parallel} = \frac{m_s}{2} v_\parallel^{\prime 2} \frac{\partial }{\partial v_\parallel^\prime}\left( v_\parallel^\prime f_s \right)
 \mathbf{b}\mathbf{b}:\nabla\mathbf{u}_s, 
    \label{eq:kpw_1X}
\end{equation}
which fully captures the dynamics of Landau damping (see Appendix~\ref{app:decomp} for a comparison $\widetilde{K}_{PS\parallel}$ and $\widetilde{K}_{PS\perp}$). Here, $\mathbf{b}$ is the unit vector along $\mathbf{B}$ so that $\mathbf{bb}:\nabla\mathbf{u}_s$ is the projection of the strain-rate tensor $\nabla\mathbf{u}_s$ along the local magnetic field. The integral of Eq.~(\ref{eq:kpw_1X}) over $v_\parallel^\prime$ is equal to the parallel pressure-strain interaction, $p_{\parallel s} \mathbf{b}\mathbf{b}:\nabla\mathbf{u}_s$. For the Langmuir wave, the magnetic field is constant along $x$ and $\mathbf{b}\mathbf{b}:\nabla\mathbf{u}$ reduces to $\partial u_x/\partial x$. For the \Alfven wave the magnetic field is not a constant, so $\mathbf{bb}:\nabla\mathbf{u}_s=b_x^2 (\partial u_{xs}/\partial x) + b_xb_y (\partial u_{ys}/\partial x) + b_xb_z (\partial u_{zs}/\partial x)$, where $b_i = B_i/B$ is the component of $\mathbf{b}$ in the $\bf{i}$-direction. Though the simulations are 1D and therefore only $x$-derivatives are present in $\mathbf{bb}:\nabla\mathbf{u}_s$, the numerical framework we employ (described in Appendix~\ref{app:model}) evolves fluid moments in all directions, which gives us access to all three spatial components of the fluid flow $\mathbf{u}_s$. 

\subsection{Standing Langmuir Wave}

Fig.~\ref{fig:kpw_fpc_slw} shows the box-integrated (a) parallel kinetic pressure-strain and (d) parallel electric field-particle correlation, given by Eq.~(\ref{eq:fpc}), as functions of velocity versus time. Velocity is normalized to the unperturbed electron thermal velocity and time to the electron plasma frequency. Panels (b) and (e) show the same quantities after applying a centered, sliding time-average with window size $\tau \omega_{pe} = 8.7$, approximately equal to two wave periods. As is well-known for the field-particle correlation, this type of time-averaging smooths out oscillations in phase-space energy density and highlights the secular transfer of energy density from fields to particles\citep{Klein:2016,Howes:2017,Klein:2017}. The phase velocities of the wave, $v_{ph}/v_{te} = \pm (\omega/\omega_{pe})/(k \lambda_{De}) = \pm1.41/0.5 = \pm2.82$, are marked by vertical dashed lines. Panels (c) and (f) contain the related fluid diagnostics in 1D, namely the pressure-strain interaction $-p_x (\partial u_x / \partial x)$ (simply the pressure dilatation) and the rate of electromagnetic work per unit volume $j_x E_x$, plotted as functions of the spatial domain normalized to the wavelength $\lambda$ versus time. Integrated over $x$, these fluid diagnostics are equivalent to the kinetic diagnostics integrated over $x$ and $v_x^\prime$, shown in (g). The box-integrated pressure-strain and the rate of electromagnetic work are approximately equal in magnitude, since the energy lost by the fields ultimately becomes electron internal energy rather than electron flow energy. We also calculate the kinetic diagnostics locally in configuration space and confirm that the value of their local integration over $v_x^\prime$ is equal to the local value of the relevant fluid quantity (not shown). However, while the fluid quantities are of comparable magnitude when averaged over the box, locally $|j_x E_x| \ll |-p_x (\partial u_x / \partial x)|$ so we choose to display the spatially-averaged diagnostics in Fig.~\ref{fig:kpw_fpc_slw}. 

\begin{figure}[t]
\includegraphics[width=0.49\textwidth,bb=20 55 515 625]{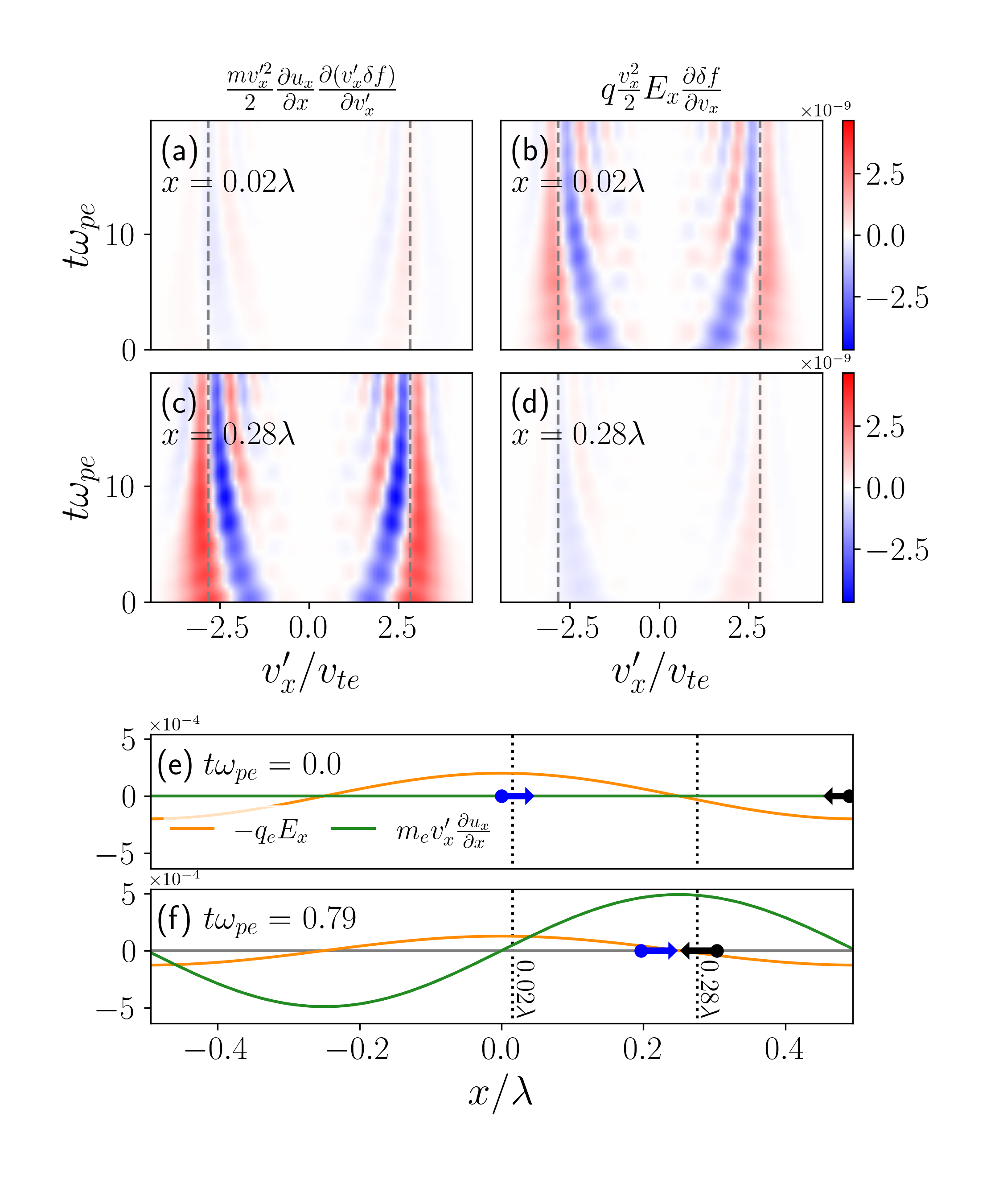}
\caption{(a), (c) The local, time-averaged, parallel KPS and (b), (d) parallel electric FPC for electrons in the Langmuir wave at two different locations: $x = 0.02\lambda$ and $x = 0.28 \lambda$. These locations are shown by vertical dotted lines in (e)-(f), which display the forces in each diagnostic, $-q_eE_x$ (orange) and $m_e v_x^\prime \partial u_x/\partial x$ for $v_x^\prime=v_{ph}$ (green), as functions of $x/\lambda$ at (e) $t=0$ and (f) $t=0.79\omega_{pe}$. The blue and black points  correspond to electrons resonantly accelerated to $|v| > |v_{ph}|$ by the electric field (arrows) at $x/\lambda=0$ (blue) and $x/\lambda=0.5$ (black) in (e) and converging at $x/\lambda \simeq 0.25$ in (f) to yield conversion of bulk kinetic to internal energy by the pressure-strain interaction.}  
\label{fig:slw_kpw_fpc_pos}
\end{figure}

To consider the kinetic diagnostics locally in configuration space, we divide the distribution function into equilibrium and fluctuating pieces, $f_s = \delta f_s + \langle f_s \rangle_x$, and use only the fluctuating portion $\delta f_s$ in order to observe the small-amplitude secular structure that is otherwise dominated by large-amplitude oscillations\citep{Klein:2016,Klein:2017,Howes:2017}. We elaborate on this subtle but important analysis change in Appendix~\ref{app:df}. For the standing wave, we find that the local KPS and FPC are dominant at different spatial locations. The top row of Fig.~\ref{fig:slw_kpw_fpc_pos} shows the (a) kinetic pressure-strain and (b) field-particle correlation at $x=0.02\lambda$, time-averaged over a centered, sliding interval $\tau \omega_{pe} = 8.7$. The second row shows the same quantities at $x=0.28\lambda$. The vertical, dotted lines in panels (e) and (f) show where these points lie in relation to the electric force $-q_e E_x$ (orange) and pressure force $m_e v_x^\prime (\partial u_x/\partial x)$ (green) of the standing wave pattern at two different times. (Note that the electric force is a constant in $v_x^\prime$, and we consider the pressure force at $v_x^\prime=v_{ph}$ in these panels.) The first point $x = 0.02\lambda$ is near a node of the pressure force, so $K_{PS\parallel}$ has a relatively small magnitude and is not symmetric about $v_x^\prime/v_{te} = 0$, as seen in panel (a). Exactly at a node of the pressure force, such as $x/\lambda = 0$, $K_{PS\parallel}$ is fully antisymmetric about $v_x^\prime/v_{te} = 0$ such that it integrates to  zero. We choose a point slightly offset from the node so that the structure of the KPS can be seen at this color scale. In contrast, the parallel electric force is near its maximum at $x = 0.02\lambda$, so the FPC has a relatively large magnitude and is nearly symmetric about $v_x^\prime/v_{te}=0$, as seen in panel (b). At $x = 0.28\lambda$ in (c) and (d), $m_e v_x^\prime \partial u_x/\partial x$ is near its maximum and $-q_e E_x$ is near zero, so that the qualitative and quantitative features of the diagnostics are reversed. 

\begin{figure*}[ht!]
\includegraphics[width=0.96\textwidth,bb=50 200 800 575]{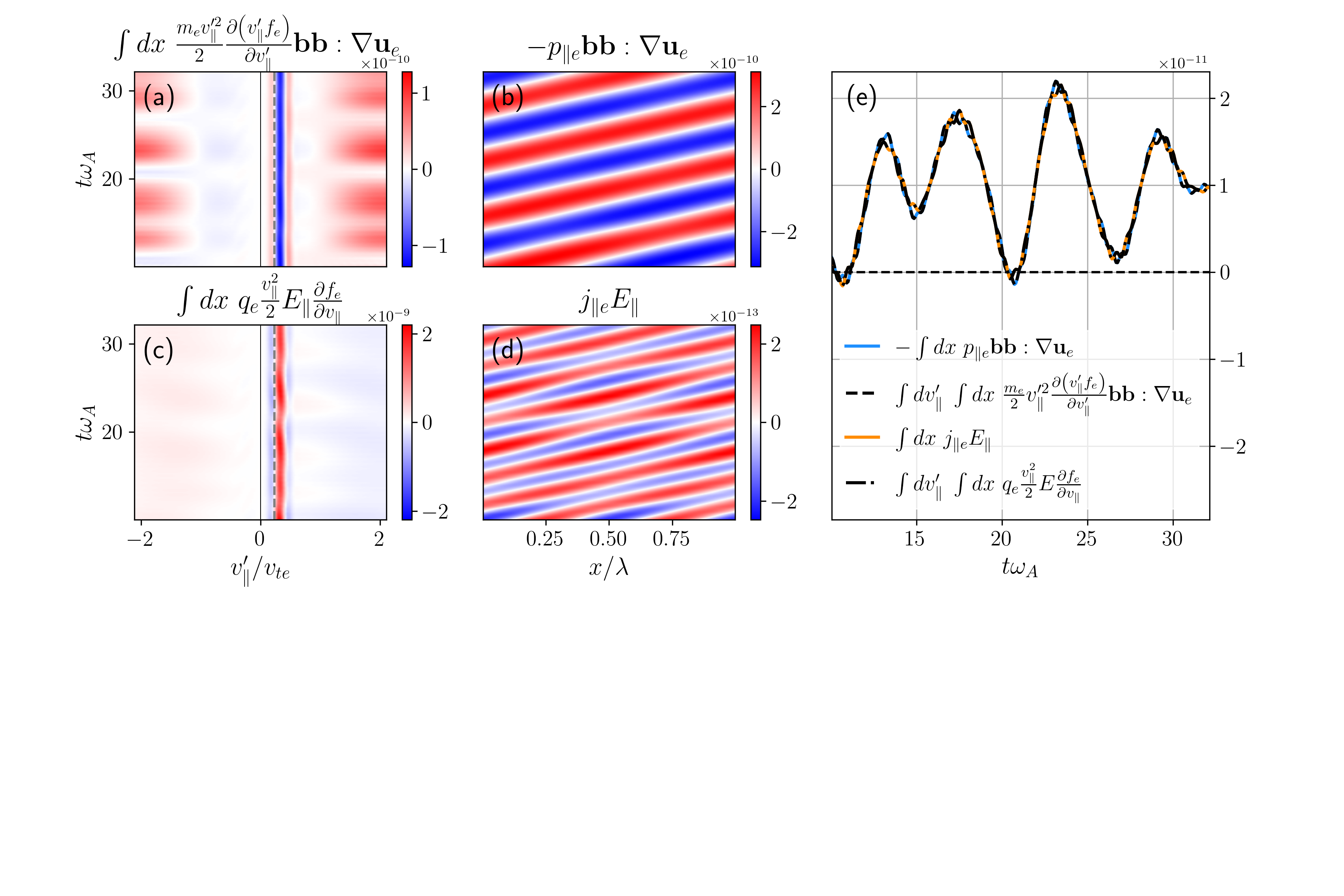}
\caption{The box-integrated (a) electron parallel KPS and (c) parallel electric FPC as functions of $v_\parallel^\prime/v_{te}$ and $t\omega_A$, and the (b) parallel fluid pressure-strain interaction and (d) parallel rate of electromagnetic work per unit volume as functions of $x/\lambda$ and $t\omega_A$ for the traveling \Alfven wave. Panel (e) shows (a) and (c) integrated over velocity and (b) and (d) integrated over space as functions of time.} 
\label{fig:aw_kpw_fpc_pos}
\end{figure*}

Physically, the bipolar FPC signature is interpreted as a net acceleration of electrons by $-q_e E_x$ from below to above the resonances at $v_x^\prime/v_{te} = \pm2.82$ as the standing wave is Landau damped. This creates an excess of particles moving just faster than the wave and a deficit of particles moving just slower than the wave\citep{Klein:2016,Horvath:2020}. The increased population of accelerated electrons stream away from $x/\lambda = 0$, as seen in (e), and converge toward the strong fluid velocity gradients at $\pm 0.28\lambda$, as seen in (f). The pressure-strain interaction in 1D is due solely to compression/expansion, so the KPS signature shows where the phase-space internal energy density of the electrons $w_e^\prime(x,v^\prime_x,t)$ is changing due to compression/expansion. Positive values in the KPS correspond to particles converging at the point in configuration space at which the diagnostic is calculated, and negative values correspond to particles diverging from that point. Therefore, the bipolar signature can be interpreted as a net increase in the phase-space internal energy density due to the convergence of resonant particles: after Landau damping produces an increased population of particles with velocities just greater than the wave phase velocity, these particles stream away from regions of strong electric force ($x= 0$ and $\pm 0.5\lambda$) into regions of strong pressure force ($x=\pm 0.25 \lambda$), where net conversion into internal energy density occurs. 

This combination of physical effects that govern the transport of energy in $(x,v)$ phase space is illustrated by the blue and black particles in Fig.~\ref{fig:slw_kpw_fpc_pos}(e) and (f).  At $t \omega_{pe}=0$ in panel (e), the blue particle at $x/\lambda =0$ represents an electron resonantly accelerated by the electric field to a velocity $v>v_{ph}$ and moving to the right (blue arrow), while the black particle at $x/\lambda =0.5$ represents an electron resonantly accelerated by the oppositely signed electric field to a velocity $v< -v_{ph}$ and moving to the left (black arrow).  After a time $\Delta t \omega_{pe}=0.79$, the blue particle with velocity $v = 3.2 v_{te} > v_{ph}$ will have moved a distance $\Delta x = v \Delta t \simeq 0.25 \lambda_{De} = 0.2 \lambda$, as shown in (f). Similarly, the black particle with velocity $v = -3.2 v_{te} < -v_{ph}$ will have moved the same distance in the opposite direction to position $x/\lambda=0.3$.  The convergence about one quarter wavelength away of these resonantly accelerated particles, depicted by the converging blue and black arrows near $x/\lambda=0.25$ in (f), about one quarter wave period after their acceleration by the electric field, leads to a conversion of their bulk flow kinetic energy into internal energy via the pressure-strain interaction. This completes the picture of the transport of energy from the electric field energy through the bulk flow kinetic energy to the internal energy in $(x,v)$ phase space.

\subsection{Traveling \Alfven Wave}
\subsubsection{Parallel Kinetic Pressure-Strain Analysis}
In Fig.~\ref{fig:aw_kpw_fpc_pos} we show the box-integrated (a) parallel kinetic pressure-strain and (c) parallel electric field-particle correlation for electrons in the \Alfven wave simulation as functions of $v_\parallel^\prime/v_{te}$ and $t\omega_A$. Note that we begin the time axis at $t\omega_A\approx10$ to avoid a spurious signal at early time, which is interpreted as fluctuations of the slow mode before it has been sufficiently damped. We begin the time axis at this value for each plot of the kinetic diagnostics for the \Alfven wave. Applying the centered, sliding time-average produces little change besides additional smoothing to the spatially-integrated diagnostics since the \Alfven wave is traveling, so we omit the time-averaged panels in this figure. Panels (b) and (d) show the related fluid diagnostics as functions of $x/\lambda$ and $t\omega_A$, and (e) shows the excellent agreement between the kinetic diagnostics integrated over space and parallel velocity and the fluid diagnostics integrated over space, as expected. The box-integrated pressure-strain (blue) and rate of electromagnetic work (orange) are nearly equal and in phase with each other, and as a result very little energy ends up in the electron fluid flow at any time. Again, along with the FPC, the KPS shows resonant, bipolar structure at the phase velocity of the Landau damped wave, $v_{ph}/v_{te} = (\omega/k_\parallel v_{ti})\sqrt{(m_e T_i/m_i T_e)} = 0.234$, which is marked with a vertical dashed line in (a) and (c). 

\begin{figure}[b!]
\includegraphics[width=0.49\textwidth,bb=15 25 580 270]
{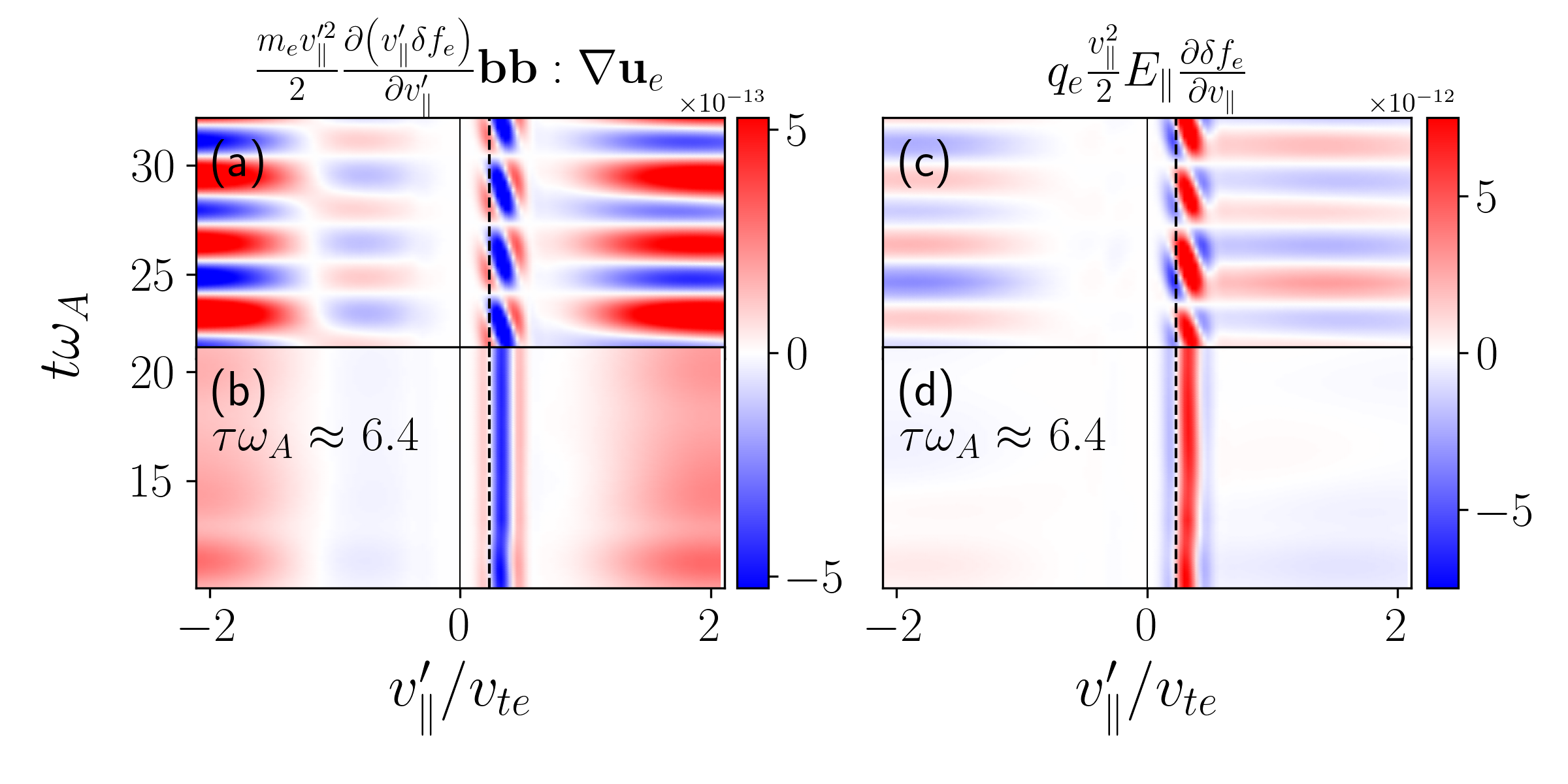}
\caption{ The (a) instantaneous ($\tau \omega_{A} = 0$) and (b) time-averaged ($\tau \omega_{A} \approx 6.4 \approx T_A $) parallel KPS and (c) instantaneous and (d) time-averaged parallel electric FPC of electrons in the traveling \Alfven wave simulation calculated at a single spatial point, $x/\lambda\approx0.6$.} 
\label{fig:aw_kpw_fpc_tau}
\end{figure}

To consider the kinetic diagnostics locally, we shift to using $\delta f_e = f_e - \langle f_e \rangle_x$ rather than the total distribution function to calculate the diagnostics. Fig.~\ref{fig:aw_kpw_fpc_tau} shows (a)-(b) the parallel kinetic pressure-strain alongside (c)-(d) the parallel field-particle correlation at a single point in configuration space. The upper panels are the instantaneous diagnostics, and the lower panels are averaged over a time interval $\tau \omega_A \approx 6.4 \approx T_A$, where $T_A=2\pi/\omega_A$ is the \Alfven wave period. The time average causes the oscillatory components of each diagnostic to cancel more completely than they did under the spatial average in Fig.~\ref{fig:aw_kpw_fpc_pos}(a) and (b), which reveals clear resonant signatures. 

Notably, Figs.~\ref{fig:aw_kpw_fpc_pos} and~\ref{fig:aw_kpw_fpc_tau} show that the resonant signatures in the kinetic pressure-strain and the field-particle correlation have bipolar structures of opposite sign, which is contrary to what was observed for the standing Langmuir wave. Mathematically, the oppositely-signed signature arises from a difference in the phase relationship between the fluctuating term in each diagnostic---$\mathbf{b}\mathbf{b}:\nabla\mathbf{u}_e$ and $E_\parallel$, respectively---and the parallel velocity perturbation to the distribution function. For either diagnostic to be strong in a given region of phase-space, these must be roughly in phase ($\phi = 0$ or $\pi$) with $\partial f_e/\partial v_\parallel$\citep{Conley:2023}. Due to the nearly instantaneous response of low-inertia electrons to changes in the electric field, the phase offset between $E_\parallel$ and $\mathbf{b}\mathbf{b}:\nabla\mathbf{u}_e$ is $\phi\approx\pi$. Therefore, the resonant diagnostics are important at the same spatial location (which changes with time as the wave propagates) but have opposite signs.

\begin{figure}[b!]
\includegraphics[width=0.49\textwidth,bb=15 25 580 290]{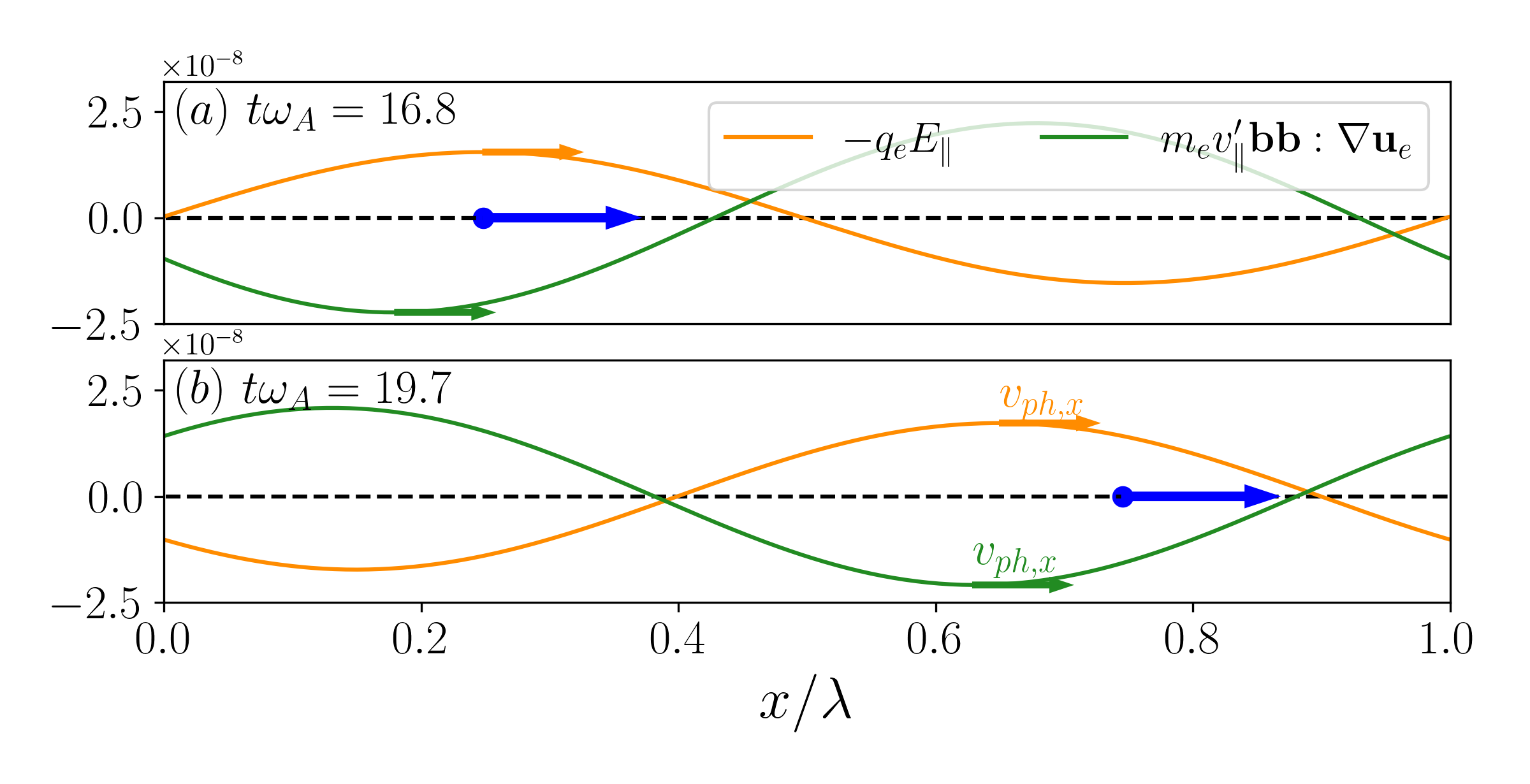}
\caption{The parallel electric (orange) and pressure
(green) forces as functions of $x/\lambda$ at (a) $t\omega_A=16.8$ and (b) $t\omega_A=19.7$. Both $-q_e E_\parallel$ and $m_e v_\parallel^\prime \mathbf{bb}:\nabla\mathbf{u}_e$ propagate to the right at $v_{ph,x}$ (maintaining a phase offset of $\phi \approx \pi$) and the accelerated, resonant particles propagate to the right at $v_x>v_{ph,x}$ (example particle in blue).} 
\label{fig:aw_diagram}
\end{figure}

\begin{figure*}[tb!]
\includegraphics[width=0.75\textwidth,bb=50 25 750 270]{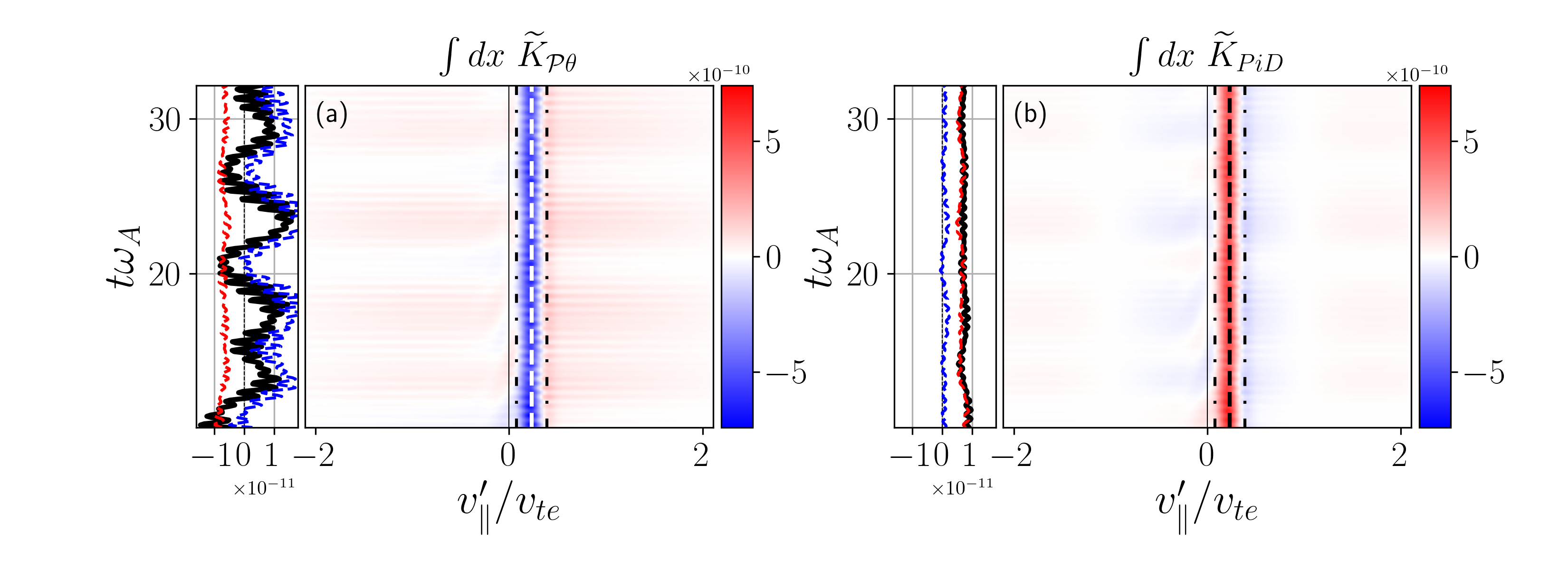}
\caption{Box-integrated electron (a) kinetic pressure dilatation and (b) kinetic Pi-D as functions of parallel velocity versus time for the \Alfven wave simulation. The panels to the left of (a) and (b) show their respective quantities integrated over all $v_\parallel^\prime$ (black), and over the resonant (red dashed) and non-resonant (blue dashed) regions of $v_\parallel^\prime$.} 
\label{fig:aw_PdivU_PiD_tau}
\end{figure*}

Physically, we understand the negative resonant signature in $K_{PS\parallel}$ in the same way we understood the positive signature in the Langmuir wave. The pressure-strain interaction is the result of compression/expansion in this geometry, so positive (negative) regions of KPS show which particles are locally converging at (diverging from) that point. Since $E_\parallel$ and $\mathbf{bb}:\nabla\mathbf{u}_e$ differ in phase by $\phi\approx\pi$ and are moving through space at speed $v_{ph}$, the particles that have been accelerated by the electric field to velocities $v>v_{ph}$ are captured in the KPS as an excess of particles diverging from that local point. This is illustrated in Fig.~\ref{fig:aw_diagram}, which compares a test particle (blue) with speed $v>v_{ph}$ to the propagation of the parallel electric force $-q_e E_\parallel$ (orange) and parallel pressure force $m_e v_\parallel^\prime \mathbf{bb}:\nabla\mathbf{u}_e$ (green), where we again evaluate the pressure force at $v_\parallel^\prime=v_{ph}$. The panels show the magnitude of these forces along $x/\lambda$ at (a) time $t\omega_A = 16.8$ and (b) time $t\omega_A = 19.7$, at which point the wave has traveled a distance of $\Delta x =  v_{ph,x} \Delta t \approx 0.4\lambda$ to the right. The phase velocity along the $x$-direction is a projection of the total, since $x$ is along $\mathbf{k}$ rather than $\mathbf{B}$, and therefore the observed speed of the wave is $v_{ph,x} \approx v_{ph}sin(-\theta)$. Note that we neglect variation in the direction of $\mathbf{B}$ in this calculation of $v_{ph,x}$, but this represents a very small correction for this linear wave. The illustrative particle has velocity $v=0.3 v_{te}>v_{ph}$, and therefore moves along $x$ at speed $v_x\approx0.3sin(-\theta)v_{te} \approx 0.028 v_{te}$. It begins (a) at $x=0.25\lambda$, near the peak of the parallel electric force. After time $\Delta t\omega_A = 2.9$ it has moved a distance of $\Delta x= v_x \Delta t \approx 0.5 \lambda$ and diverged with respect to the phase of the traveling wave, as shown in (b).

\subsubsection{Kinetic Analogs of Pressure Dilatation and Pi-D}
When the KPS is integrated over all velocity space, we find a net positive value of $-(\mathbf{P}_e\cdot \nabla)\cdot \mathbf{u}_e$, both locally and upon integrating over configuration space. (Note that velocity space is restricted to $\pm2.1v_{te}$ in these figures to highlight the resonant structure. See Appendix~\ref{app:decomp} for figures with velocity-space extents of $\pm4.5v_{te}$). A positive, box-integrated pressure-strain indicates net conversion of energy from $E^f_e$ to $E^{int}_e$. This implies that the observed increase of $E^{int}_e$ over the course of the simulation shown in Fig.~\ref{fig:aw_energybalance}(a) comes from non-resonant electrons, while the resonant electrons contribute a decrease to the internal energy density as the \Alfven wave is damped. 

To help interpret this counter-intuitive result, we decompose the pressure-strain interaction into the pressure dilatation and Pi-D (as in Eq.~(\ref{eq:pdivu_PiD})) and consider kinetic analogs of these terms. Though the simulation box is 1D, we capture dynamics perpendicular to the magnetic field through a spectral expansion, and therefore have access to all components of the fluid velocity $\mathbf{u}_s$ and the pressure tensor $\mathbf{P}_s$, which results in non-zero Pi-D. In terms of parallel and perpendicular velocity, the kinetic electron pressure dilatation is defined here by
\begin{equation}
    \widetilde{K}_{\mathcal{P}\theta} \equiv -\frac{1}{3} m_ef_e \left( v_\parallel^{\prime 2} + v_\perp^{\prime 2} \right)\nabla \cdot \mathbf{u}_e,
    \label{eq:kpTh}
\end{equation}
and kinetic electron Pi-D by 
\begin{equation}
    \widetilde{K}_{PiD} \equiv -m_e f_e \left( v_\parallel^{\prime 2} - \frac{v_\perp^{\prime 2}}{2}\right) \left( \mathbf{bb}:\nabla\mathbf{u}_e - \frac{1}{3} \nabla \cdot \mathbf{u}_e \right).
\label{eq:kPiD}
\end{equation}
When added together, these two expressions equal the alternative kinetic pressure-strain $\widetilde{K}_{PS}$, given in Eq.~(\ref{eq:akpw}). We choose to write the kinetic decomposition in terms of the alternative form of the KPS since the decomposition into pressure dilatation and Pi-D relates directly to the fluid form of the pressure-strain interaction, like $\widetilde{K}_{PS}$ does. Despite not having $v^{\prime}_\perp$ as a coordinate in our simulation, $v_\perp^{\prime 2} f_e/2$ is equivalent to a portion of the spectral representation of the distribution function (called $\mathcal{G}$, see Appendix~\ref{app:model}) that is retained in the simulation, which allows us to compute these terms. The general forms of Eqs.~(\ref{eq:kpTh}) and~(\ref{eq:kPiD}) are included in Appendix~\ref{app:decomp}. In 1D, both the compressible (pressure dilatation) and incompressible (Pi-D) contributions to the full pressure-strain interaction come solely from converging/diverging flows. The velocity-space structure in $\widetilde{K}_{PS}$ measures changes in $\partial w^\prime/\partial t$ that directly correspond to the pressure-strain interaction. Therefore, features in $\widetilde{K}_{\mathcal{P}\theta}$ and $\widetilde{K}_{PiD}$ indicate regions of phase-space which contain particles that are converging to (positive values) or diverging from (negative values) a spatial point, and provide a net contribution to the pressure-strain. Note that for Landau damping, resonant structures in the alternative KPS are centered at $v_{ph}$ rather than organized in a bipolar signature about $v_{ph}$ (see Fig.~\ref{fig:akps_par_perp} in Appendix~\ref{app:decomp}).

Fig.~\ref{fig:aw_PdivU_PiD_tau} shows (a) the box-integrated kinetic pressure dilatation and (b) box-integrated kinetic Pi-D. The resonant signature in $\int dx\ \widetilde{K}_{\mathcal{P}\theta}$ is negative, while the resonant signature in $\int dx\ \widetilde{K}_{PiD}$ is positive. Integrating these quantities over all $v_\parallel^\prime$ to obtain the box-integrated fluid quantities (black lines in the left panels of (a) and (b), respectively), we see that $-\mathcal{P}_e\theta_e$ oscillates about zero but has a net positive value while $-\mathbf{\Pi}_e:\mathbf{D}_e$ is always positive. The contribution of the resonant signatures is isolated by defining a resonant range of parallel velocity, $0.08 \leq \Delta v_{\parallel R}^\prime/v_{te} \leq 0.39$, where the extent of $\Delta v_{\parallel R}^\prime/v_{te}$ is determined by the width of the signature in the alternative FPC (not shown). The resonant region is bounded by vertical dash-dotted black lines in Fig.~\ref{fig:aw_PdivU_PiD_tau}. As expected, the velocity integral over $\Delta v_{\parallel R}^\prime$ (red dashed lines in left-hand panels) is negative for the kinetic pressure dilatation and positive for kinetic Pi-D. The integral over the remainder of velocity space (blue dashed lines) is positive and oscillating for the pressure dilatation and approximately zero for Pi-D. Thus, this decomposition of $\widetilde{K}_{PS}$ for the \Alfven wave shows that both its resonant and non-resonant features are dominated by the pressure dilatation. (Note that this is qualitatively true for $K_{PS\parallel}$, shown in Figs.~\ref{fig:aw_kpw_fpc_pos}(a) and~\ref{fig:aw_kpw_fpc_tau}(b), and quantitatively true for $\widetilde{K}_{PS}$, plotted in Fig.~\ref{fig:akps_par_perp}(c), Appendix~\ref{app:decomp}). However, if we consider the cumulative contributions of the pressure dilatation and Pi-D, we find that Pi-D---and therefore the resonant particles---plays a major role in the conversion from fluid flow to internal energy in the process of Landau damping.

\begin{figure}[h!]
\includegraphics[width=0.49\textwidth,bb=0 10 430 250]{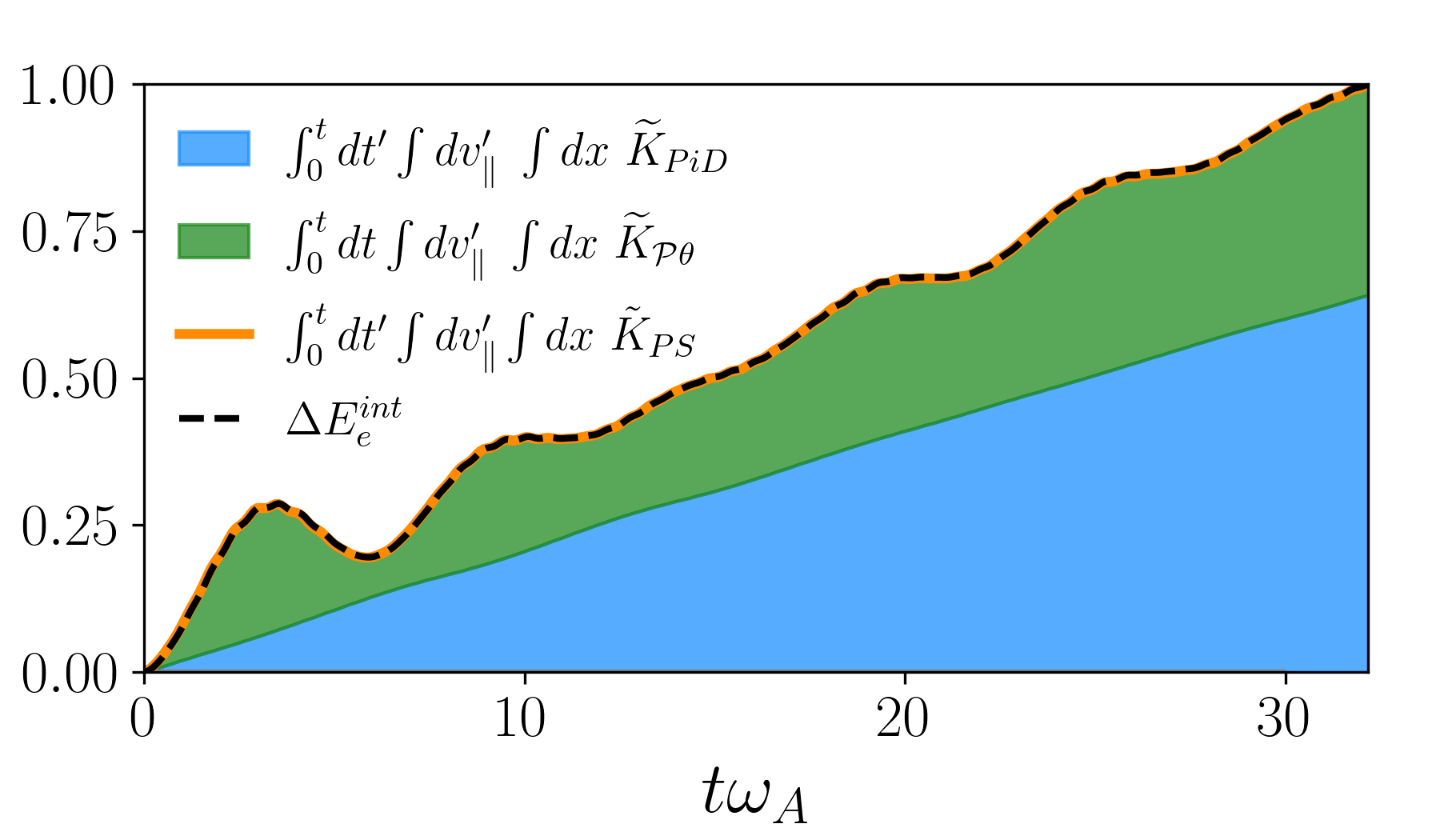}
\caption{Box-integrated change in electron internal energy $\Delta E_e^{int}$ (black dashed) and cumulative value of the box-integrated alternative KPS integrated over velocity (orange); the cumulative contributions of box-integrated Pi-D (blue) and pressure dilatation (green), calculated from their kinetic analogs.} 
\label{fig:aw_kpw_PdivU_PiD_cum}
\end{figure}

Fig.~\ref{fig:aw_kpw_PdivU_PiD_cum} shows the cumulative contributions over time of the box-integrated electron pressure dilatation, $\int_0^t dt^\prime \int dv_\parallel^\prime\ \int dx\ \widetilde{K}_{\mathcal{P}\theta}$ (green fill), and box-integrated electron Pi-D, $\int_0^t dt^\prime \int dv_\parallel^\prime \int dx\ \widetilde{K}_{PiD}$ (blue fill). Together, these equal the cumulative contribution of the box-integrated electron pressure-strain interaction, $\int_0^t dt^\prime \int dv_\parallel^\prime\ \int dx\ \widetilde{K}_{PS}$ (orange line) and the total net change in electron internal energy $\Delta E_e^{int}$ (black dashed). All values are normalized to the magnitude of $\Delta E_e^{int}$ at time $t\omega_A=32.2$. Though the sign of the resonant and non-resonant features in the KPS indicate that the increase in electron internal energy is due entirely to the non-resonant particles, the separate cumulative contributions of the pressure dilatation and Pi-D show that the total energy conversion from $E^f_e$ to $E^{int}_e$ is dominated by incompressible effects. Further, the kinetic analog of Pi-D shows that this positive contribution is almost entirely due to the incompressible convergence of resonant particles. 

\section{Summary and Conclusion}\label{sec:disc}
In this work we present a new tool for phase-space analysis of energy transport in collisionless plasmas: the Kinetic Pressure-Strain (KPS). Just as the field-particle correlation technique\citep{Klein:2016,Howes:2017,Klein:2017} provides a valuable velocity-space perspective on transfer between electromagnetic field energy and particle kinetic energy---which enables physical energization mechanisms to be identified---the kinetic pressure-strain provides a velocity-space view of energy conversion between bulk flows and internal energy for its fluid counterpart, the pressure-strain interaction $-\mathbf{P} \cdot (\nabla \cdot \mathbf{u})$. In both case studies of Landau damping by electrons that we consider here, the kinetic pressure-strain $K_{PS}$ contains a prominent bipolar velocity-space structure organized around the phase velocity of the damped wave, just as the field-particle correlation does. This indicates the importance of the Landau resonance in driving not only energy transfer between fields and particles, but also energy conversion between fluid flow and internal energy. Further, it demonstrates that this new diagnostic can provide insight into physical energization processes via velocity space signatures much like the established field-particle correlation technique. 

In the standing Langmuir wave, we find that the resonant signatures of electron Landau damping look very similar in the FPC and the KPS, but occur at different times and in different spatial locations. This shows that in certain physical scenarios the KPS plays a complementary role to the FPC in identifying spatially localized mechanisms of energy transfer, although the FPC is necessary in order to ensure that the total energy transfer from fields to particles is understood. The signature in the FPC indicates that the resonant particles are first responsible for positive transfer from electromagnetic energy density $\mathcal{E}^m$ to electron fluid flow energy density $\mathcal{E}_e^f$ as the wave experiences Landau damping. This generates an elevated population of resonant particles with velocities just above $v_{ph}$ at points of elevated electric force, $-q_eE_x$. The fluid flow energy increases temporarily as the particles stream away from these points toward regions of elevated pressure force, $m_ev_x^\prime \partial u_x/\partial x$. There, strong resonant signatures in the KPS are observed which indicate positive conversion of energy from $\mathcal{E}_e^f$ to electron internal energy density $\mathcal{E}_e^{int}$. Physically, the increase in internal energy occurs as the excess population of resonant particles converges at the point of strong flow gradient.

For the traveling \Alfven wave, we again find that the resonant signature of electron Landau damping in the KPS looks very similar to the resonant signature in the FPC, but in this case having the opposite sign. Since electrons have a nearly instantaneous response to the electric field, $E_\parallel$ and $\mathbf{bb}:\nabla\mathbf{u}_e$ differ in phase by $\phi \approx \pi$, and the oppositely-signed signature arises because these quantities have opposite phase relationships with $\partial f_e/\partial v_\parallel^\prime$. Physically, the signature in the KPS is understood as a divergence of the increased population of resonant particles away from the location of the extreme value of $\mathbf{bb}:\nabla\mathbf{u}_e$. At approximately the same spatial point as this extreme value, the electric field creates an enhanced population of resonant particles traveling faster than $v_{ph}$ (an increase in phase-space kinetic energy density). This population of particles diverges with respect to $E_\parallel$ and $\mathbf{bb}:\nabla\mathbf{u}_e$ which travel at $v_{ph}$, and therefore a negative signature is created in the KPS. The negative sign of this signature indicates that the resonant particles participate in energy conversion from $E_e^{int}$ to $E_e^{f}$, and the non-resonant particles are responsible for the overall increase in internal energy as the \Alfven wave damps. This is in contrast to what we observed in the case of the Langmuir wave, where resonant particles were responsible for the energy transport at each step.

When the pressure-strain interaction is separated into the pressure dilatation and Pi-D, however, a different understanding emerges for the role of the resonant particles. The cumulative contribution of box-integrated Pi-D is larger than the cumulative contribution of the box-integrated pressure-dilatation. Further, the kinetic analog to this term, $\widetilde{K}_{PiD}$, shows that the magnitude of Pi-D is due almost entirely to particles near the resonant velocity of the damped \Alfven wave. Therefore, incompressible convergence of resonant particles is responsible for the majority of the conversion from fluid flow to internal energy as the \Alfven wave in this simulation experiences Landau damping. This underscores the value decomposing the pressure-strain interaction in order to properly parse the physics of collisionless particle energization that it contains\citep{Yang:2017a,Cassak:2022a}, and also highlights the additional value brought to the table by considering kinetic analogs of these energy transport diagnostics. 

This introduction of the kinetic pressure-strain as a new diagnostic of collisionless energy transport in phase-space illustrates how it may be used to understand kinetic physics that is necessarily obscured by studying fluid diagnostics alone. Additionally, we present kinetic analogs to the decomposition of the pressure-strain interaction into pressure dilatation and Pi-D for the traveling \Alfven wave simulation. Studying these diagnostics in the expanded detail of full phase-space further elucidates the physics of the pressure-strain interaction and conversion between fluid flow and internal energy. Further study of the kinetic pressure-strain, its kinetic decompositons, and the field-particle correlation together in a variety of physical scenarios will develop a more complete understanding of how these diagnostics can be effectively leveraged together to further our understanding of collisionless plasma energization. 

\section*{Acknowledgements}
The simulation framework described in Appendix B and the numerics will be published in forthcoming papers. The theory for this model was jointly developed by J.~Juno, J.~TenBarge and A.~Hakim (PPPL), and the numerics by J.~Juno and A.~Hakim. Part of this work is summarized in Section IV.A and Appendix~\ref{app:model}. We acknowledge the contributions of A.~Hakim on these aspects of the work presented here. We also acknowledge helpful conversations with M.~A.~Shay and H.~Perera.

S.~A.~Conley was supported by a NSF Atmospheric and Geospace Science Postdoctoral Fellowship, Grant No.~AGS-2318252. J.~Juno was supported by the U.S. Department of Energy under Contract No. DE-AC02-09CH1146 via an LDRD grant. J.~M.~TenBarge was supported by NASA grant 80NSSC23K0099. P.~A.~Cassak and M.~H.~Barbhuiya were supported by NASA grant 80NSSC24K0172, and P.~A.~Cassak was additionally supported by Department of Energy grant DE-SC0020294. G.~G.~Howes was supported by NSF grant AGS-1842561 and NASA grants 80NSSC24K0260 and 80NSSC24K0552. E.~Lichko was supported by NSF grant 1949802. J.~Juno and J.~M.~TenBarge acknowledge the Frontera computing project at the Texas Advanced Computing Center. Frontera is made possible by National Science Foundation (NSF) award OAC-1818253. 

\section*{Declaration of interests}
The authors have no conflicts to disclose.

\newpage 
\appendix 
\section{Lab-Frame Derivation of the Kinetic Pressure-Strain}
\label{sec:labframe}

Here, we derive an expression for the conversion of energy between bulk and internal phase space energy densities in a stationary (lab) reference frame.  We begin from the collisional Vlasov equation in conservative form,
\begin{equation}
    \frac{\partial f_s}{\partial t} +  \nabla \cdot \left(\mathbf{v} f_s \right) + \nabla_{v} \cdot \left[ \frac{q_s}{m_s} \left(\mathbf{E} + \mathbf{v} \times \mathbf{B}\right)  f_s \right]=C_s(f),
    \label{eq:labframeVlasov}
\end{equation}
where $C_s(f)$ is the collision operator. Multiplying by $(1/2)m_s v^2$ and introducing $w_s = (1/2)m_s v^2 f_s$ gives
\begin{equation}
\frac{\partial w_s}{\partial t} + \nabla \cdot (\mathbf{v} w_s) + \frac{1}{2} q_s v^2 \left( \mathbf{E} + \mathbf{v} \times \mathbf{B} \right) \cdot \nabla_v f_s = \frac{1}{2} m_s v^2 C_s(f).
\end{equation}
The third term on the left hand side contains the standard field-particle correlation term, and the right hand side describes phase space energy density conversion via collisions.  Focusing on the second term on the left hand side, we write $\mathbf{v} = \mathbf{u}_s + \mathbf{v}^\prime$, so the term becomes
\begin{eqnarray}
\nabla \cdot (\mathbf{v} w_s) & = & \nabla \cdot \left( \frac{1}{2} m_s v^2 \mathbf{v} f_s \right) \\
& = & \nabla \cdot \left( \frac{1}{2} m_s u_s^2 \mathbf{u}_s f_s \right) + \nabla \cdot \left( m_s (\mathbf{u}_s \cdot \mathbf{v}^\prime) \mathbf{u}_s f_s \frac{}{} \right) \nonumber \\  & + & \nabla \cdot \left(\frac{1}{2} m_s v^{\prime 2} \mathbf{u}_s f_s \right) \label{eq:labframews} + \nabla \cdot \left( \frac{1}{2} m_s u_s^2 \mathbf{v}^\prime f_s \right) \\ & + & \nabla \cdot \left( m_s (\mathbf{u}_s \cdot \mathbf{v}^\prime) \mathbf{v}^\prime) f_s \frac{}{} \right) + \nabla \cdot \left(\frac{1}{2} m_s v^{\prime 2} \mathbf{v}^\prime f_s \right). \nonumber
\end{eqnarray}
The six terms in Eq.~(\ref{eq:labframews}) represent the six ways that a flux of phase space energy density can locally change the phase space energy density $w_s$.

To make a connection with the known fluid description of energy evolution, we simply integrate Eq.~(\ref{eq:labframews}) over all velocity space.  In particular, the integral of the first term is the flux of bulk kinetic energy density, $\nabla \cdot [(1/2) m_s n_s u_s^2 \mathbf{u}_s] = \nabla \cdot (\mathcal{E}^k_s \mathbf{u}_s)$.  The velocity space integral of the second term and fourth terms vanish identically because the average of $\mathbf{v}_s^\prime$ vanishes.  The velocity space integral of the third term is $\nabla \cdot [(3/2) \mathbf{u}_s P_s] = \nabla \cdot (\mathcal{E}^{int}_s \mathbf{u}_s)$, where $P_s = \int (1/3) m_s v^{\prime 2} f_s d^3v$ is the scalar pressure (one third of the trace of the pressure tensor).  
Skipping to the velocity space integral of the sixth term, it is the divergence of the heat flux density vector $\nabla \cdot \mathbf{q}$, where $\mathbf{q} = \int (1/2) m_s v^{\prime 2} \mathbf{v}^\prime f_s$.  Finally, before taking the velocity space integral of the fifth term, we write it using index notation as 
\begin{equation}
\frac{\partial}{\partial x_j} (m_s u_{sk} v^\prime_{k} v^\prime_{j} f_s) = \frac{\partial u_{sk}}{\partial x_j} (m_s v^\prime_{k} v^\prime_{j} f_s) + u_{sk} \frac{\partial}{\partial x_j} (m_s v^\prime_{k} v^\prime_{j} f_s). \label{eq:pstermplus}
\end{equation}
The velocity space integral of the first of these two terms is $(\partial u_{sk}/\partial x_j) P_{s,jk} = (\mathbf{P}_{s} \cdot \nabla) \cdot \mathbf{u}_s$, which is the negative of the pressure-strain interaction.  The first term in Eq.~(\ref{eq:pstermplus}) is therefore equivalent to $\widetilde{K}_{PS}$ in Eq.~(\ref{eq:akpw}), but here obtained from the lab frame perspective.  The velocity space integral of the other term gives $\mathbf{u}_s \cdot (\nabla \cdot \mathbf{P}_s)$, which is the rate of work per unit volume done by the pressure force.  Thus, $\nabla \cdot (\mathbf{v} w_s)$ describes the phase space energy density of conversion between bulk and internal energy, but resolved as a function of phase space. It contains the phase space density associated with the pressure-strain interaction.  We leave to future work a more in-depth analysis of the other terms.

\section{Hybrid Discretization Model}
\label{app:model}
The model utilized in this manuscript is a newly implemented hybrid discretization of the Vlasov-Maxwell system based on the current discontinuous Galerkin (DG) implementation in \gke, combined with an optimized spectral basis in a subset of the degrees of freedom. 
Consider the Vlasov equation in the local flow frame,
\begin{equation*}
\begin{split}
    \frac{\partial f}{\partial t} & + \nabla \cdot \left[\left(\mathbf{v}^\prime + \mathbf{u}\right) f \right] \\
    & + \nabla_{v^\prime} \cdot \left[ \left( \frac{1}{\rho} \nabla \cdot \mathbf{P}\ + \frac{q}{m} \mathbf{v}^\prime \times \mathbf{B} - \mathbf{v}^\prime \cdot \nabla\mathbf{u} \right) f \right] = 0, \\
\end{split}
\end{equation*}
where we have dropped the species subscript $s$ for notational simplicity.
We make a further transformation to the Vlasov equation in this form to utilize a velocity-space coordinate system aligned with the local magnetic field, i.e., 
\begin{align}
  \mvec{v}'
  & =
  \mvec{v}'(\mvec{v}'',\mvec{x}'',t'')
  =
  v_\parallel^\prime \buni(\mvec{x}',t) + \mvec{v}_\perp^\prime (\mvec{x}',t) \notag \\
  & =
  v_\parallel^\prime \buni(\mvec{x}',t) +
  v_\perp^\prime\cos\theta^\prime \gvec{\tau}_1(\mvec{x}',t) + v_\perp^\prime\sin\theta^\prime \gvec{\tau}_2(\mvec{x}',t),
\end{align}
where $\mvec{v}'' = v_\parallel^\prime \buni + \mvec{v}_\perp^\prime$ are our new velocity coordinates, $\buni$ is the local magnetic field unit vector and $\gvec{\tau}_1, \gvec{\tau}_2$ define the plane perpendicular to the local magnetic field, and as before the spatial and temporal coordinates are unchanged by this transformation, $\mvec{x}'' = \mvec{x}', t'' = t'$.
Because the local magnetic field unit vector $\buni$ and the plane perpendicular to the local magnetic field, spanned by $\gvec{\tau}_1, \gvec{\tau}_2$, all depend on space and time, the Vlasov equation further changes to include the forces arising from using this velocity-space coordinate system. 
For completeness, we quote the Vlasov equation in this coordinate system, 
\begin{align}
  \pfrac{\bdf}{t}
  +
  \delxpp\cdot\left[ (v_\parallel^\prime \buni + \mvec{v}_\perp^\prime+\mvec{u})\bdf \right]
  +
  \frac{1}{v_\perp^\prime}\frac{\partial}{\partial w^i}
  \left( v_\perp^\prime a^i \bdf \right)
  = 0,
  \label{eq:vlasov-cgl}  
\end{align}
where now $\bdf = f(\mvec{x}'',v_\parallel^\prime,v_\perp^\prime,\theta^\prime,t)$, $w^i = (w^1, w^2, w^3) = (v_\parallel^\prime, v_\perp^\prime, \theta^\prime)$ are our new velocity coordinates, and
$a^i = \tdbasis{i}\cdot\mvec{a}''$ are the components of the
acceleration in this frame of reference moving with the local flow and aligned with the local magnetic field.
To determine the components of the acceleration in this frame of reference, we dot the acceleration vector
\begin{align}
  \mvec{a}'' =
  \pfrac{\mvec{v}''}{t}
  +
  (\mvec{v}''+\mvec{u})\cdot\delx\mvec{v}''
  +
  \frac{1}{\rho}\delx\cdot\mvec{P}
  -\mvec{v}''\cdot\delx\mvec{u}
  +
  \frac{q}{m} \mvec{v}''\times\mvec{B}.
\end{align}
with the duals of the tangent vectors $\tdbasis{i}$, which are given by first computing the tangent vectors as
\begin{align}
  \tbasis{i} = \pfrac{\mvec{v}'}{w^i},
\end{align}
and then solving for the duals with either finding the inverse of the metric for this coordinate transformation $g_{ij} = \tbasis{i} \cdot \tbasis{j}$, $g^{ij} = \left ( g_{ij} \right )^{-1}$, $\tdbasis{j} = g^{ij} \tbasis{i}$ or, in this simple transformation, identifying the dual subject to the constraint $\tbasis{i} \cdot \tdbasis{j} = \delta_i^j$.
In this case, the tangent vectors are $\tbasis{\parallel} = \buni$,
\begin{align}
  \tbasis{\perp} &= \cos\theta^\prime \gvec{\tau}_1 + \sin\theta^\prime \gvec{\tau}_2, \\
  \tbasis{\theta} &= -v_\perp^\prime \sin\theta^\prime \gvec{\tau}_1 + v_\perp^\prime \cos\theta^\prime \gvec{\tau}_2,
\end{align}
and thus the dual basis $\tdbasis{i}$ vectors are
$\tdbasis{\parallel} = \tbasis{\parallel}$,
$\tdbasis{\perp} = \tbasis{\perp}$ and
\begin{align}
  \tdbasis{\theta} &= -\frac{1}{v_\perp^\prime} \sin\theta^\prime \gvec{\tau}_1
                    + \frac{1}{v_\perp^\prime} \cos\theta^\prime \gvec{\tau}_2.
\end{align}
The Jacobian of the transform, $\mathcal{J}_v = \sqrt{\det(g_{ij})}$ is simply $\mathcal{J}_v = v_\perp^\prime$. In terms of $\tbasis{\perp}$ we can write $\mvec{v}_\perp^\prime =
v_\perp^\prime \tbasis{\perp}$.

We emphasize no information has been lost by performing these transformations. 
\eqr{\ref{eq:vlasov-cgl}} still conserves all the same invariants of the Vlasov equation and still obeys phase-space incompressibility. What we gain is the ability to represent the different degrees of freedom of this collisionless plasma with different numerical methods, and thus can optimize our resolution in the case of magnetized plasmas where the dynamics parallel and perpendicular to the magnetic field separate. 

In particular, motivated by the fact that a collisionless, magnetized plasma's motions are restricted perpendicular versus parallel to the field, as well as recent successes in spectral velocity discretizations of the Vlasov equation\citep{Delzanno:2015, Parker:2015, Vencels:2016, Roytershteyn:2018, Koshkarov:2021, Pagliantini:2023}, we expand the distribution function perpendicular to the field in a Fourier-Laguerre basis: 
\begin{align}
    f(\mvec{x}, \mvec{v}'', t) =  \sum_{k=0}^{\infty}  \sum_{\ell=-\infty}^{\infty} & \mathcal{F}_{k\ell}(\mvec{x}, v_\parallel^\prime, t)\exp \left (-\frac{m v_\perp^{\prime2}}{2 T_\perp (\mvec{x}, t)}  \right ) \notag \\
    & \times \frac{m}{2 \pi T_\perp(\mvec{x}, t)} L_k \left (\frac{ m v_\perp^{\prime2}}{2 T_\perp(\mvec{x}, t)} \right ) e^{i\ell\theta^\prime}. \label{eq:pkpm_f}
\end{align}
Here, $L_k$ are the Laguerre polynomials,
\begin{align}
    L_m(x) = \frac{e^x}{m!} \frac{d^m}{dx^m} \left (e^{-x} x^m \right ),
\end{align}
$e^{i\ell\theta^\prime}$ are the Fourier basis, and $\mathcal{F}_{k\ell}(\mvec{x}, v_\parallel^\prime, t)$ are the spectral coefficients encoding the representation of the distribution function in this spectral expansion.
We discretize the resulting $\mathcal{F}_{k\ell}(\mvec{x}, v_\parallel^\prime, t)$ equations with a DG method for however many Laguerre polynomials and Fourier harmonics we wish to retain, i.e., however much perpendicular velocity-space resolution we require.
Further details of the theoretical foundation and numerical implementation of this hybrid discretization, which we call a ``parallel-kinetic-perpendicular-moment'' model, will be detailed in a forthcoming publication. 

For now, we emphasize that this approach works well for the studies described in this manuscript, where the plasma remains mostly gyrotropic and thus can be represented with a small amount of perpendicular velocity-space resolution, only a single Fourier harmonic and two Laguerre polynomials. 
In this limit, we solve two kinetic equations, 
\begin{align}
  &\pfrac{F_{0}}{t}
  +
  \delx\cdot\left[ (v_\parallel^\prime\buni+\mvec{u})F_{0} \right] \notag \\
  &+
  \frac{\partial}{\partial v_\parallel^\prime}
  \left[
  \buni\cdot
  \left(
  \frac{1}{\rho}\delx\cdot\mvec{P} - v_\parallel^\prime\buni\cdot\delx\mvec{u}
  \right)  F_{0}
  + \mathcal{G}\delx\cdot\buni
  \right] = 0, \\
  &\pfrac{\mathcal{G}}{t}
  +
  \delx\cdot\left[ (v_\parallel^\prime\buni+\mvec{u})\mathcal{G} \right] \notag \\
  &+
  \frac{\partial}{\partial v_\parallel^\prime}
  \left[
  \buni\cdot
  \left(
  \frac{1}{\rho}\delx\cdot\mvec{P} - v_\parallel^\prime\buni\cdot\delx\mvec{u}
  \right)  \mathcal{G} 
  +
  \left(
  4\mathcal{G} - 2 \frac{T_\perp}{m} F_0
 \right)\frac{T_\perp}{m}\delx\cdot\buni
  \right] \notag \\
  &=
  \left[
  \buni\cdot(\buni\cdot\delx\mvec{u})
  - \delx\cdot(v_\parallel^\prime\buni + \mvec{u})
  \right] \mathcal{G}.  
\end{align}
Here the full distribution function, $f(\mvec{x}, \mvec{v}'', t)$ is related to the quantities in these two kinetic equations as
\begin{align}
    \mathcal{F}_{00}(\mvec{x}, v_\parallel^\prime, t) & = F_0(\mvec{x}, v_\parallel^\prime, t), \\
    \mathcal{F}_{10}(\mvec{x}, v_\parallel^\prime, t) & = F_0(\mvec{x}, v_\parallel^\prime, t) - \frac{m}{T_\perp(\mvec{x}, t)} \mathcal{G}(\mvec{x}, v_\parallel^\prime, t), 
\end{align}
which are then substituted into \eqr{\ref{eq:pkpm_f}} to reconstruct the full phase space variation of the truncated distribution function for the subsequent analysis in phase space.
These two kinetic equations are coupled to the conservation of momentum equation to determine the flow velocity, 
\begin{align}
    \pfrac{(\rho \mvec{u})}{t} + \delx\cdot\left(\rho \mvec{u} \mvec{u} + \mvec{P} \right) = \rho \frac{q}{m} \left( \mvec{E} + \mvec{u} \times \mvec{B} \right),
\end{align}
plus Maxwell's equations for the evolution of the electromagnetic fields,
\begin{align}
  \frac{\partial \mvec{B}}{\partial t} + \delx\times\mvec{E} &= 0, \label{eq:dbdt} \\
  \epsilon_0\mu_0\frac{\partial \mvec{E}}{\partial t} - \delx\times\mvec{B} &= -\mu_0\mvec{J}, \label{eq:dedt} \\
  \delx\cdot\mvec{E}&= \frac{\varrho_c}{\epsilon_0}, \label{eq:divE} \\
  \delx\cdot\mvec{B}&= 0. \label{eq:divB}
\end{align}
with the coupling to the electromagnetic fields arising directly from the momentum equation through the plasma currents,
\begin{align}
    \mvec{J} = \sum_s \frac{q_s}{m_s} \rho_s \mvec{u}_s.
\end{align}
We note that the pressure tensor $\mvec{P}$ is obtained via velocity moments of the two distribution function coefficients that are evolved in time,
\begin{align}
    p_\parallel & = m \int v_\parallel^{\prime2} F_0 \thinspace dv_\parallel^\prime, \\
    p_\perp & = m \int \mathcal{G} \thinspace dv_\parallel^\prime, \\
    \mvec{P} & = (p_\parallel - p_\perp) \buni \buni + p_\perp \mvec{I}.
\end{align}
The full parallel dynamics, including collisionless Landau damping from parallel electric fields, are contained in this ``parallel-kinetic-perpendicular-moment'' model, and it is straight-forward to extract both the classical field-particle correlation energization in the lab frame, as well as the pressure-strain interactions parallel to the field, and examine both of these diagnostics in phase space. 

\section{The Perturbed Distribution Function}
\label{app:df}

Here we elaborate on the importance of the shift from using the total distribution function $f_s$ to using the perturbed distribution function $\delta f_s = f_s - \langle f_s \rangle_x$ in our analysis of the KPS and FPC at a single point in space. As was originally discussed regarding the field-particle correlation technique\citep{Klein:2016,Klein:2017,Howes:2017}, it is necessary to consider only the perturbed portion of the distribution function in order to isolate net changes in the phase-space energy density $w_s(\mathbf{x}, \mathbf{v}, t)$ when computing the FPC. Using the full distribution function retains large-magnitude, conservative oscillations in $\partial w_s/\partial t$ that overwhelm the small-amplitude, net changes to the phase-space energy density. Thus, the signatures of net energy transfer that are created by various energization mechanisms (e.g., the resonant bipolar signature of Landau damping) are masked by using full $f_s$. The same principle holds for observing net changes in the phase-space internal energy density $w_s^\prime(\mathbf{x}, \mathbf{v}^\prime, t$) in the KPS. 

\begin{figure}[b!]
\includegraphics[width=0.49\textwidth,bb=30 20 600 280]{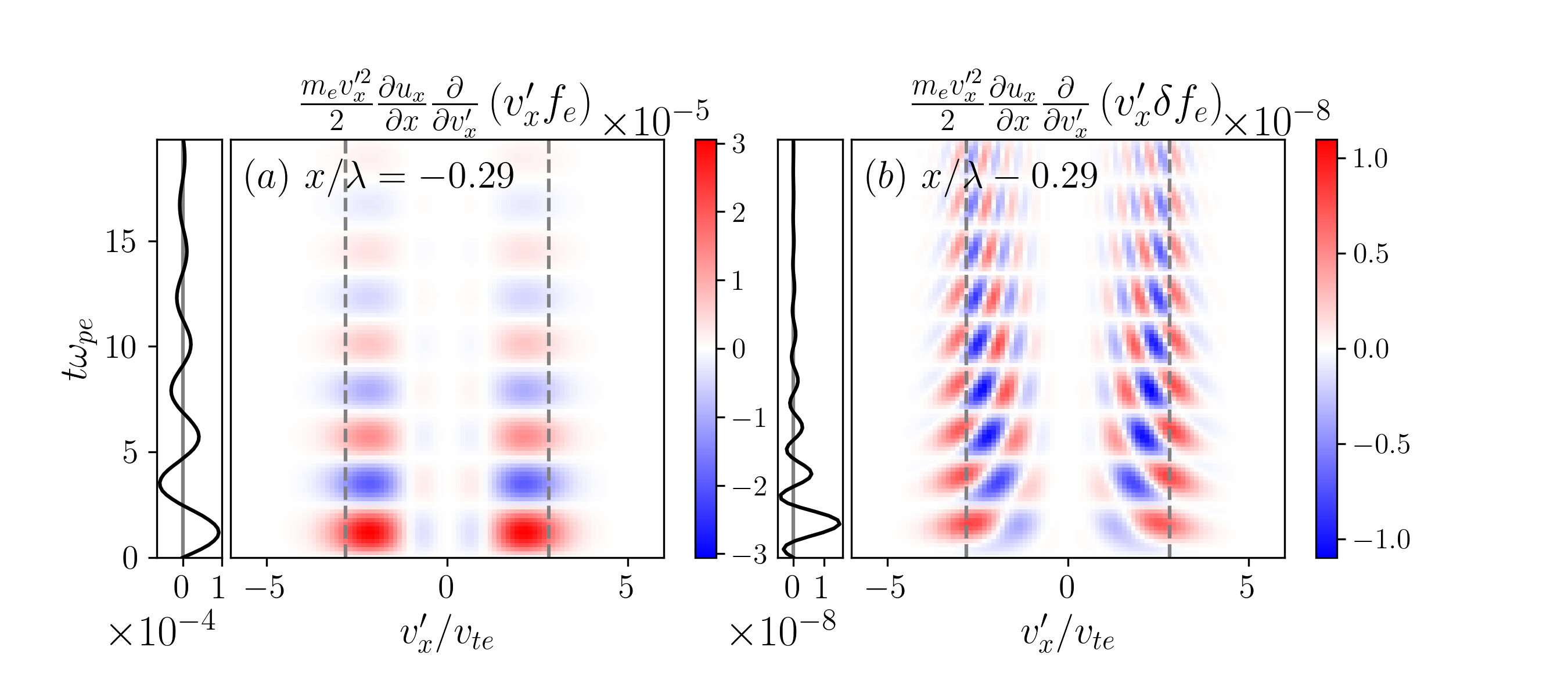}
\caption{$K_{PS}$ calculated locally at $x/\lambda=-0.29$ using (a) the total distribution function $f_e$ and (b) the perturbed distribution function $\delta f_e$ in the Langmuir wave simulation. The left-hand panels show (a) and (b) integrated over velocity.} 
\label{fig:slw_f_df}
\end{figure}

If the diagnostic is integrated over the entire spatial domain (such as in Figs.~\ref{fig:kpw_fpc_slw}  and~\ref{fig:aw_kpw_fpc_pos}), it is possible to see the resonant signatures using total $f_s$ because the average over a full wave-period cancels the oscillating components of the energy transfer. However, in order to see the resonant signatures locally, the perturbed distribution $\delta f_s$ must be used. This is shown in Fig.~\ref{fig:slw_f_df}, which compares the local KPS calculated using (a) $f_e$ and (b) $\delta f_e = f_e - \langle f_e \rangle_x$ in the standing Langmuir wave simulation. Note that the structure in (b), which is organized around the phase-velocity of the wave $v_{ph}=\pm 2.82 v_{te}$ (grey dashed lines), is three orders of magnitude smaller than the structure in (a). The large-magnitude features of the full $f_e$ KPS are even in velocity-space, so that when the KPS is integrated over $v_x^\prime$ (left panel), we see that the local pressure-strain interaction is also dominated by large-magnitude oscillations. 

For our results, is it necessary to calculate the perturbed portion of the distribution function using a quasilinear-like approach, where we separate the distribution into an equilibrium and a perturbed portion, $f_s = f_{s0} + \delta f_s$, and calculate the equilibrium distribution at each output time using a spatial average over the simulation box, $f_{s0} = \langle f_s \rangle_x$. 

\section{Decomposition of the Kinetic Pressure-Strain}
\label{app:decomp}
In this work we consider kinetic forms of the decomposition of the pressure-strain interaction into pressure dilatation and Pi-D, 
\begin{equation}
    -(\mathbf{P}_s \cdot \nabla)\cdot \mathbf{u}_s = -\mathcal{P}_s\theta_s - \mathbf{\Pi}_s : \mathbf{D}_s.
\end{equation}
The general, 3D-3V kinetic forms of this decomposition are defined by
\begin{equation}
    \widetilde{K}_{\mathcal{P}\theta} \equiv -\frac{1}{3} m_s v^{\prime  2} f_s \left(\nabla \cdot \mathbf{u}_s \right)
    \label{eq:3D3V_kpTh}
\end{equation}
and
\begin{equation}
    \widetilde{K}_{PiD}
 \equiv -m_s \left[ \left( \mathbf{v}^\prime \mathbf{v}^\prime - \frac{1}{3}v^{\prime 2}\mathbf{I} \right) f_s\right] : \mathbf{D}_s.
 \label{eq:3D3V_kPiD}
\end{equation}
These kinetic forms are related to the alternative kinetic pressure-strain, $\widetilde{K}_{PS}$. Indeed, it is easy to verify that adding Eqs.~(\ref{eq:3D3V_kPiD}) and (\ref{eq:3D3V_kpTh}) produces Eq.~(\ref{eq:akpw}).  

\begin{figure*}[htb!]
\includegraphics[width=0.98\textwidth,bb=60 40 700 415]{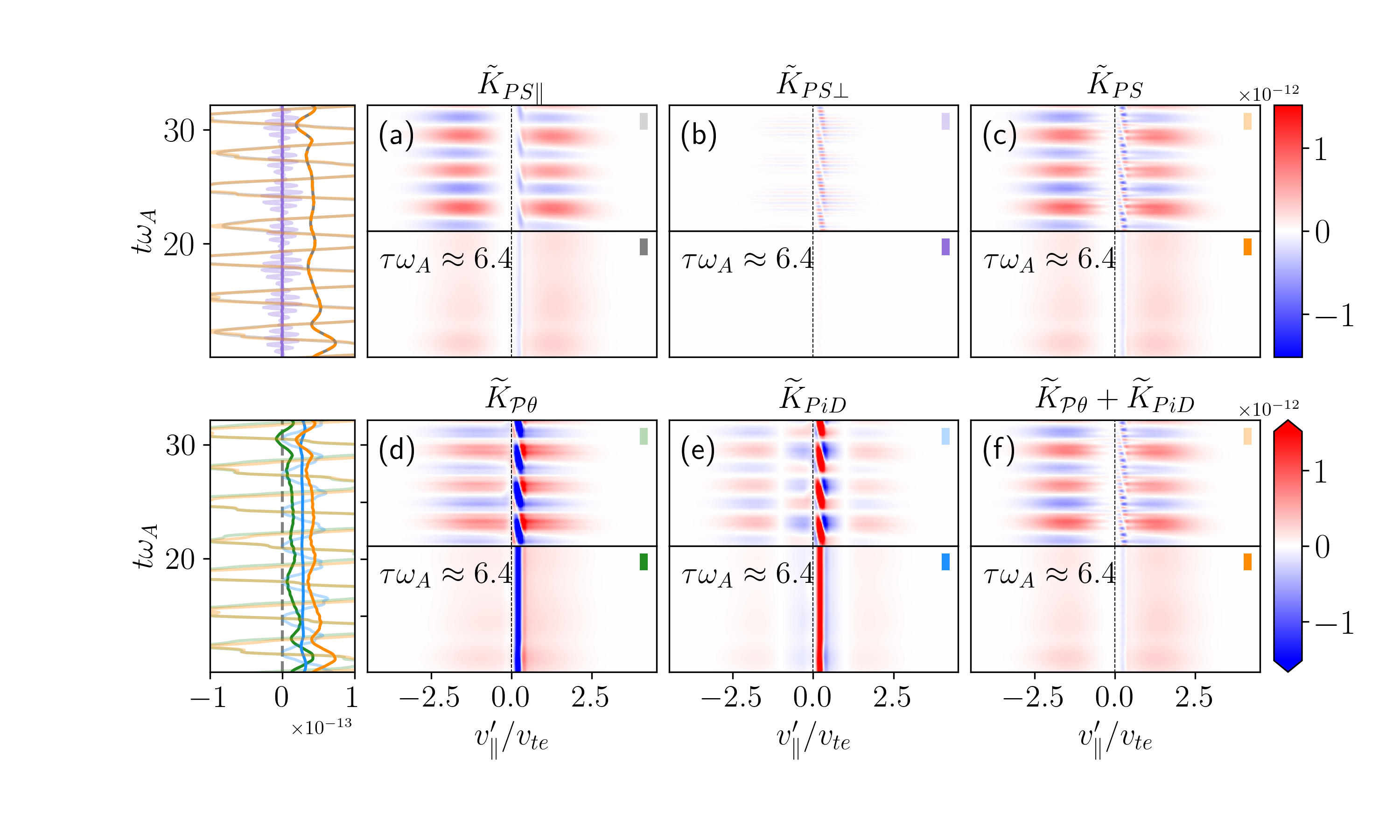}
\caption{Comparison of the (a) parallel and (b) perpendicular components of (c) the alternative kinetic pressure-strain, along with (d) the kinetic pressure dilatation and (e) kinetic Pi-D in the \Alfven wave simulation. (f) The sum of the kinetic pressure dilatation and kinetic Pi-D. Upper panels are instantaneous values, and lower panels are time-averaged with a centered, sliding window of $\tau\omega_A\approx6.4$. All diagnostics are calculated locally at $x/\lambda \approx 0.6$ Left-hand panels show the quantities in each row integrated over $v_\parallel^\prime/v_{te}$; the colored rectangles in the upper right of each panel provide the legend. Note that the color-scale is the same for all plots, and the lower row contains over-saturated values.} 
\label{fig:akps_par_perp}
\end{figure*}

In a standard Vlasov-Maxwell simulation in 1D-1V, the kinetic pressure dilatation reduces to $-m_s v_x^{\prime 2} f_s (\partial u_{xs} / \partial x)$ (which is Eq.~(\ref{eq:akpw}) in 1D-1V), and the kinetic Pi-D is zero. In our simulation model, perpendicular moments such as flow velocity $\mathbf{u}_s$ and pressure $\mathbf{P}_s$ are included, so despite having only a single kinetic spatial and velocity coordinate, the truly 1D-1V forms of Eqs.~(\ref{eq:3D3V_kPiD}) and (\ref{eq:3D3V_kpTh}) do not contain the full amount of information to which we have access. The plamsa is gyrotropic, so it is natural to write $\mathbf{v}^\prime$ in terms of $v_\parallel^\prime$ and $v_\perp^\prime$. Then, Eq.~(\ref{eq:3D3V_kpTh}) becomes
\begin{equation}
    \widetilde{K}_{\mathcal{P}\theta} = -\frac{1}{3} m_s f_s (v_\parallel^{\prime 2} + v_\perp^{\prime 2}) \left( \nabla \cdot \mathbf{u}_s \right),
    \label{eq:1D_kpTh}
\end{equation}
which is the expression given in Eq.~(\ref{eq:kpTh}), and Eq.~(\ref{eq:3D3V_kPiD}) becomes
\begin{equation}
\begin{aligned}
     \widetilde{K}_{PiD} = -m_s f_s 
     		\left[ 
		v_\parallel^{\prime 2} \mathbf{bb} + \frac{v_\perp^{\prime 2}}{2}  \left( \mathbf{I} - \mathbf{bb} \right) \right. \enspace &
		\\
     		\left. -\frac{1}{3} \left(v_\parallel^{\prime 2} + v_\perp^{\prime 2} \right)\mathbf{I} \right] & : \mathbf{D}_s. 
\end{aligned}
\end{equation}
The double contractions with $\mathbf{D}_s$ simplify to $\mathbf{bb}:\mathbf{D}_s = \mathbf{bb}:\nabla\mathbf{u}_s - (1/3)\nabla \cdot \mathbf{u}_s$, $(\mathbf{I} - \mathbf{bb}):\mathbf{D}_s = -\mathbf{bb}:\nabla\mathbf{u}_s + (1/3)\nabla \cdot \mathbf{u}_s$, and $\mathbf{I}:\mathbf{D}_s = 0$ since $\mathbf{D}_s$ is traceless. Kinetic Pi-D is therefore 
\begin{equation}
    \widetilde{K}_{PiD} = -m_s f_s \left( v_\parallel^{\prime 2} - \frac{v_\perp^{\prime 2}}{2} \right)\left( \mathbf{bb}:\nabla\mathbf{u}_s - \frac{1}{3}\nabla \cdot \mathbf{u}_s \right),
    \label{eq:1D_kPiD}
\end{equation}
which is Eq.~(\ref{eq:kPiD}). Note that these equations are general for any number of spatial coordinates (assuming gyrotropy). 

Finally, we emphasize that when Eqs.~(\ref{eq:1D_kpTh}) and (\ref{eq:1D_kPiD}) are added together, the result is the full alternative kinetic pressure-strain, Eq.~(\ref{eq:akpw}). This is easy to see by dividing $\widetilde{K}_{PS}$ into parallel and perpendicular components, $\widetilde{K}_{PS} = \widetilde{K}_{PS\parallel} + \widetilde{K}_{PS\perp}$, where 
\begin{equation}
    \widetilde{K}_{PS\parallel}= -m_s f_s v_\parallel^{\prime 2} \mathbf{bb}:\nabla \mathbf{u}_s
\end{equation}
is the parallel alternative kinetic pressure-strain and
\begin{equation}
    \widetilde{K}_{PS\perp} = -m_s f_s \frac{v_\perp^{\prime 2}}{2} \left( \nabla \cdot \mathbf{u}_s - \mathbf{bb}:\nabla\mathbf{u}_s \right)
\end{equation}
is the alternative perpendicular kinetic pressure-strain. 

Fig.~\ref{fig:akps_par_perp} compares (a) the parallel alternative kinetic pressure-strain $\widetilde{K}_{PS\parallel}$, (b) the perpendicular alternative kinetic pressure-strain $\widetilde{K}_{PS\perp}$ and (c) the full $\widetilde{K}_{PS}$. The top panels are the instantaneous diagnostics, and the bottom panels have been time-averaged with a centered, sliding window of length $\tau\omega_A\approx6.4$. The unlettered panels on the left show the quantities integrated over $v_\parallel^\prime$, where the legend is provided by the colored rectangles. The first row illustrates that the physics of Landau damping is fully captured by the parallel KPS. The integral over velocity of the time-averaged $\widetilde{K}_{PS\parallel}$ (dark grey) is equal to that of $\widetilde{K}_{PS}$ (dark orange), while the integral over velocity of the time-averaged $\widetilde{K}_{PS\perp}$ (dark purple) is negligible. The second row of this figure shows (d) the kinetic pressure dilatation $\widetilde{K}_{\mathcal{P}\theta}$, (e) kinetic Pi-D $\widetilde{K}_{PiD}$, and (f) their sum $\widetilde{K}_{\mathcal{P}\theta} + \widetilde{K}_{PiD}$, which is equal to $\widetilde{K}_{PS}$. This shows the large magnitude of the resonant signatures in  $\widetilde{K}_{\mathcal{P}\theta}$ and $\widetilde{K}_{PiD}$ compared with the resonant signal in $\widetilde{K}_{PS\parallel}$, which is comparable to that in $\widetilde{K}_{PS}$. In the velocity-integrated panel for this row, we see that locally, the magnitude of time-averaged Pi-D (dark blue) is also larger than than of the time-averaged pressure dilatation (dark green). 

\newpage
%

\end{document}